\documentclass[12pt, a4paper]{article}
\pdfoutput=1

\usepackage[table]{xcolor}
\usepackage{amsmath}
\usepackage{amsfonts}
\usepackage{amssymb}
\usepackage{graphicx, rotating}
\usepackage{epstopdf}
\usepackage{epsfig}
\usepackage{latexsym}
\usepackage{graphicx}
\usepackage{color}
\usepackage{amsmath,amssymb}
\usepackage{cite}
\usepackage{bbold}
\usepackage{yfonts}
\usepackage{slashed}
\usepackage{hyperref}
\hypersetup{colorlinks, citecolor=bluscuro, linkcolor=black, urlcolor=bluscuro}
\definecolor{rossos}{cmyk}{0,1,1,0.55}
\definecolor{bluscuro}{rgb}{0.15, 0.2, .85}
\definecolor{bluchiaro}{cmyk}{1,.3,0.,0.1}
\usepackage{cancel}
\def\hhref#1{\href{http://arxiv.org/abs/#1}{#1}} 
\definecolor{oucrimsonred}{rgb}{0.6, 0.0, 0.0}
\definecolor{persianblue}{rgb}{0.11, 0.22, 0.73}
\definecolor{forestgreen}{rgb}{0.13,0.35,0.13}
 \hypersetup{colorlinks, citecolor=oucrimsonred, linkcolor=persianblue, urlcolor=oucrimsonred}
 \hypersetup{colorlinks, citecolor=oucrimsonred, linkcolor=persianblue, urlcolor=oucrimsonred}

\usepackage{lmodern}

\makeatletter
\ifcase \@ptsize \relax
  \newcommand{\miniscule}{\@setfontsize\miniscule{4}{5}}
\or
  \newcommand{\miniscule}{\@setfontsize\miniscule{5}{6}}
\or
  \newcommand{\miniscule}{\@setfontsize\miniscule{5}{6}}
\fi
\makeatother

\newcommand{\heart}{\ensuremath\heartsuit}

\usepackage{lipsum}

\newcommand{\be}{\begin{equation}}
\newcommand{\ee}{\end{equation}}
\newcommand{\bea}{\begin{eqnarray}}
\newcommand{\eea}{\end{eqnarray}}
\newcommand{\bc}{\begin{center}}
\newcommand{\ec}{\end{center}}

\usepackage{adjustbox}
\usepackage{array}
\usepackage{booktabs}
\usepackage{multirow}

\newcolumntype{R}[2]{%
    >{\adjustbox{angle=#1,lap=\width-(#2)}\bgroup}%
    l%
    <{\egroup}%
}

\def\Re{{\rm Re\,}}

 \def\Ph{\Phi}

\begin{document}


\vspace*{-2cm}
\begin{flushright}
CERN-TH-2017-245 \\
IPPP/17/88\\
\end{flushright}

\begin{center}
\vspace*{15mm}

\vspace{1cm}
{\Large \bf 
Black hole superradiance \\\vspace{.2cm}
and polarization-dependent bending of light
} \\
\vspace{1.4cm}

\bigskip

{\bf Alexis D. Plascencia$^a$ and Alfredo Urbano$^{b,c}$}
 \\[5mm]

{\it $^a$  Institute for Particle Physics Phenomenology, Department of Physics, Durham University, \\[1mm]
     Durham DH1 3LE, United Kingdom.}\\[1mm]
{\it $^b$ INFN, sezione di Trieste, SISSA, via Bonomea 265, 34136 Trieste, Italy.}\\[1mm]
{\it $^c$ CERN, Theoretical Physics Department, Geneva, Switzerland.}\\[1mm]

\end{center}
\vspace*{10mm} 
\begin{abstract}\noindent\normalsize
An inhomogeneous pseudo-scalar field configuration behaves like an optically active medium. 
Consequently, if a light ray passes through an  axion cloud surrounding a Kerr black hole, it may
  experience a polarization-dependent bending.
We explore the size and relevance of such effect considering both the QCD axion and a generic axion-like particle.

\end{abstract}

\vspace*{3mm}


\pagebreak

\tableofcontents

\pagebreak

\section{Motivation}

Superradiance is a radiation enhancement process which occurs  in the presence of a dissipative system.
 We refer the interested reader to~\cite{Brito:2015oca} for an excellent and comprehensive review 
 about the role of superradiance in astrophysics and particle physics.
 In the following, we
  highlight  the main aspects that are relevant for our analysis.
  
 In General Relativity, black hole superradiance is permitted in the case of Kerr black holes 
 by the presence of the event horizon and the
  ergoregion~\cite{Press:1972zz,Press:1973zz,Teukolsky:1974yv,Damour:1976kh,Zouros:1979iw,Detweiler:1980uk}. 
 The former is, for all intents and purposes, a one-way viscous
membrane from which nothing, at least at the classical level, can escape. 
In other words, the presence of an event horizon makes  black holes perfect absorbers. The latter is a region surrounding the event horizon where everything -- literarily, including light -- is forced to co-rotate with the black hole.
The presence of both the event horizon and the ergoregion 
creates the ideal conditions to make the Penrose process -- that is the extraction of energy from a rotating black hole -- possible~\cite{Penrose:1971uk}. Black hole superradiance can be thought of as the wave analogue of the Penrose process.

Superradiance has remarkable consequences in the presence of a confining 
mechanism, for instance provided by the presence of a perfectly 
reflecting mirror surrounding the black hole. In this case 
the amplified pulse bounces back and forth, 
exponentially increasing its amplitude, and eventually leading to an instability. 
This situation is naturally realized when the Kerr black hole is coupled to a massive boson since 
 low-frequency radiation is confined due to a Yukawa-like suppression.
 
 Let us make these points more quantitative following the same line of reasoning 
  presented in ~\cite{Brito:2015oca,Detweiler:1980uk}.
 We consider a massive wave-packet  in the gravitational field of a black hole.
The situation is remarkably similar to that of an electron in the Coulomb potential of 
an  hydrogen atom, and the problem -- after introducing the 
tortoise coordinate $r^*$, with $r^* \to -\infty$ as $r$ approaches the black hole horizon $r_+$ -- reduces to the solution of a Schr\"odinger-like one-dimensional equation $d^2\Psi/dr^{* 2} + V_{\rm eff}(r)\Psi = 0$ describing the radial motion under the influence of an effective potential. 
 For a Schwarzschild black hole of mass $M$ the effective potential takes the form
 \begin{equation}\label{eq:PotSchw}
 V_{\rm eff}^{\rm Schw}(r) =  \omega^2 - \left(
 1 - \frac{2G_N M}{r}
 \right)\left[
 \frac{2G_N M}{r^3} + \frac{l(l+1)}{r^2} + \mu^2
 \right]~,
 \end{equation}
 where
 $G_N = (1/M_{\rm Pl})^2$ is the Newton's constant (with $M_{\rm Pl} \simeq 1.22\times 10^{19}$ GeV the Planck mass),
  $\omega$ is the frequency of the wave-packet, $\mu$ the scalar field mass, and $l$ the 
 azimuthal  quantum number. The structure of eq.~(\ref{eq:PotSchw}) remarks the analogy with the 
 hydrogen atom mentioned before with a gravitational potential -- instead of the usual Coulomb
  contribution -- 
 besides the centrifugal term. 
Asymptotically, considering both the horizon at $r\to r_+$ (equivalently, $r^*\to -\infty$) and spatially infinity at $r\to \infty$, the most general solution is 
 \begin{equation}\label{eq:GeneralSystem}
\Psi \sim
\left\{
\begin{array}{c}
  \mathcal{T}e^{-ik_+r^*} + \mathcal{O}e^{ik_+ r^*}~~~~~ r \to r_+~,\\
  \mathcal{R}e^{ik_{\infty}r^*} + \mathcal{I}e^{-ik_{\infty} r^*}~~~~~ r \to \infty~,
\end{array}
\right.
\end{equation}
with $k_+^2 \equiv V_{\rm eff}(r \to r_+)$, $k_{\infty}^2 \equiv V_{\rm eff}(r \to {\infty})$, and 
generic transmitted ($\mathcal{T}$), reflected ($\mathcal{R}$), incident
 ($\mathcal{I}$), and outgoing ($\mathcal{O}$) flux. 
 In the following simplified discussion we assume the potential to be real even if 
 this is not true in general because $\omega$ is a complex number.
 Since under this assumption the Schr\"odinger equation is real, the complex conjugate of any
solution is also a solution. We can, therefore, impose the Wronskian equality
 $\left.\mathcal{W}(\Psi,\Psi^*)\right|_{r\to r_+} = \left.\mathcal{W}(\Psi,\Psi^*)\right|_{r\to \infty}$, with $\mathcal{W}(\Psi,\Psi^*) \equiv \Psi(\Psi^*)^{\prime}-\Psi^*(\Psi)^{\prime}$, and we find 
 the unitarity condition~\cite{Brito:2015oca}
 \begin{equation}\label{eq:Reflected}
 |\mathcal{R}|^2 = |\mathcal{I}|^2 - \frac{k_{+}}{k_{\infty}}
 \left(
 |\mathcal{T}|^2 - |\mathcal{O}|^2
 \right)~.
 \end{equation}
 Notice that for
  a black hole at the horizon the outgoing flux at the horizon is zero, $\mathcal{O} = 0$, at least at the classical level.
 The wave is superradiant amplified, i.e. $|\mathcal{R}|^2 > |\mathcal{I}|^2$, if $k_{+}/k_{\infty} < 0$. 
 For the Schwarzschild black hole in eq.~(\ref{eq:PotSchw}) one immediately finds that 
 $\left.k_{+}/k_{\infty}\right|_{\rm Schw} = \omega/\sqrt{\omega^2 - \mu^2}$, and the 
 superradiant condition never happens.
 On the contrary, since $\mathcal{O} = 0$, we find  $|\mathcal{R}|^2 < |\mathcal{I}|^2$ that is the typical case of an absorber material.
  Let us now move to the case of a Kerr black hole with mass $M$ and angular momentum $J = aM$. 
 The effective potential is more complicated (see eqs.~(\ref{eq:KGTortoise},\,\ref{eq:GWPotential}) below) but it is straightforward to find
 \begin{equation}
 \left.\frac{k_{+}}{k_{\infty}}\right|_{\rm Kerr} = \left(\omega - \frac{a m}{2G_N Mr_+}\right)/\sqrt{\omega^2 - \mu^2}~.
 \end{equation}
 The superradiant condition is verified if $\omega < am/2G_N Mr_+$, where $-l \leqslant m \leqslant l$ is the 
 magnetic quantum number, and the reflected wave is superradiant 
  amplified.
  
 This simple example makes clear the general features of black hole superradiance outlined at the beginning of the section. 
 First of all, the importance of the horizon. In the absence of an horizon -- consider for instance a generic star -- it is necessary to impose 
 a regularity condition at the center. As a consequence of $d\Psi/dr|_{r\to 0}=0$,
 the Wronskian at the center vanishes. The Wronskian at infinity 
 gives $\left.\mathcal{W}(\Psi,\Psi^*)\right|_{r\to \infty} = -2ik_{\infty}(|\mathcal{R}|^2 - |\mathcal{I}|^2)=0$, and there is no superradiance since $|\mathcal{R}|^2 = |\mathcal{I}|^2$. 
 More generally, this is the typical condition that occurs in the absence of a dissipative mechanism 
 because in this case conservation of energy implies that 
 the outgoing flux equals
the transmitted one, $|\mathcal{T}|^2 = |\mathcal{O}|^2$, and the condition $|\mathcal{R}|^2 = |\mathcal{I}|^2$ follows from eq.~(\ref{eq:Reflected}).\footnote{In the absence of an horizon, superradiance is possible only in the presence of an alternative dissipation mechanism.
 See~\cite{Cardoso:2017kgn} for an interesting recent example in the context of conducting rotating stars.}
 Second, we see that the black hole spin $a\neq 0$ is crucial to fulfill the superradiant condition, and rotational 
 energy powers the growth of the reflected wave in eq.~(\ref{eq:Reflected}).
The extraction is made possible because the rotational energy of a Kerr black hole is not located 
inside the event horizon but in the ergoregion.
 This is the crucial difference compared to the Schwarzschild case, in which there is no energy available outside the event horizon. Finally, the presence of a mass term $\mu$ naturally provides a confining mechanism 
 for the low-frequency reflected waves since if $\omega <\mu$ from $e^{ik_{\infty}r}$ and $k_{\infty} = \sqrt{\omega^2 - \mu^2}$ 
 one gets a Yukawa-like suppression.
 
 The striking conclusion that follows from this discussion is that, under the specific conditions that trigger a superradiant instability, in the presence of a massive
  scalar field it should not be possible to observe fast-spinning black holes simply because the black hole
  must spin down as a consequence of energy extraction.\footnote{Superradiance is also possible for a massive spin-1~\cite{Pani:2012vp,Baryakhtar:2017ngi} and spin-2 field~\cite{Brito:2013wya}.}
 Black hole spin-measurements~\cite{McClintock:2013vwa,Reynolds:2013qqa} are therefore a valid experimental observable to constraint or discover new massive scalar particles~\cite{Arvanitaki:2010sy}.
 As a rule of thumb, superradiance is relevant if the Compton wavelength of the massive particle
 $\lambda_{\rm Compton} = 1/\mu$
  is of the same order compared with the black hole radius $R \approx 2G_{N}M$
 \begin{equation}
 M \approx 6.7\,\left(
 \frac{10^{-12}\,{\rm eV}}{\mu}
 \right)\,M_{\odot}~.
 \end{equation}
Supermassive black hole with $M \sim 10^6$ $M_{\odot}$  corresponds to ultra-light scalar with $\mu \sim 10^{-18}$ eV while stellar-mass black holes are relevant if $\mu \sim 10^{-12}$ eV. 
  From a particle physics perspective, 
  such light scalars are natural if protected by some underlying symmetry that makes the presence of 
  a tiny mass term technically natural, and 
  the most convincing case is that of a psuedo-Nambu--Goldstone boson, a light scalar field arising from the spontaneous breaking of a global symmetry. 
  The QCD axion and, more generally, axion-like particles (ALPs) are typical examples. The former is theoretically motivated by the solution of the strong CP problem, the latter are 
  ubiquitous in the low-energy limit of string constructions  (the ``axiverse''~\cite{Arvanitaki:2009fg}).
   Black hole superradiance is, therefore, an extremely interesting discovery tool for this class of new physics particles.
   
 However, the story told so far only relies on gravitational interactions. 
 In other words, any boson with mass $\mu$, irrespective of its particle physics origin, 
 will display the same physics as far as the aforementioned picture of superradiance 
 is concerned. 
 
 The goal of this paper is to present and discuss an observable 
 consequence of black hole superradiance that is intimately connected to the axionic nature of the scalar cloud. To this end, we shall exploit the axion effective coupling to photons which is 
 defined by the Lagrangian density 
\begin{equation}\label{eq:AxionPhotonL}
\mathcal{L}_{a\gamma\gamma} = \frac{g_{a\gamma\gamma}}{4}\Phi F_{\mu\nu}\tilde{F}^{\mu\nu} = 
-g_{a\gamma\gamma}\Phi \vec{E}\cdot\vec{B}~.
\end{equation}
In the case of the QCD axion this coupling -- inherited from the mixing 
with light mesons ($\pi_0$, $\eta$, $\eta^{\prime}$, {\it et cetera}) as 
well as by the triangle anomaly of the Peccei-Quinn fermions -- is in general non vanishing and it motivates 
a rich search strategy based on axion-photon conversion
in external magnetic fields~\cite{Carosi:2013rla}.

Our idea is very simple, and can be illustrated as follows. 
Consider an electromagnetic wave in the vacuum, defined by the wave vector
 $\hat{k}  = \vec{k}/|\vec{k}|$ determining the direction of propagation, the angular frequency $\omega$, and  two basis polarization vectors $\hat{e}_{i=1,2}$, both being perpendicular 
 to $\hat{k}$. Under parity, we have the transformation property 
 $(\hat{k}, \hat{e}_1, \hat{e}_2) \overset{P}{\to} (\hat{k}, \hat{e}_1, -\hat{e}_2)$. 
 \begin{figure}[!htb!]
\begin{center}
  \includegraphics[width=.85\linewidth]{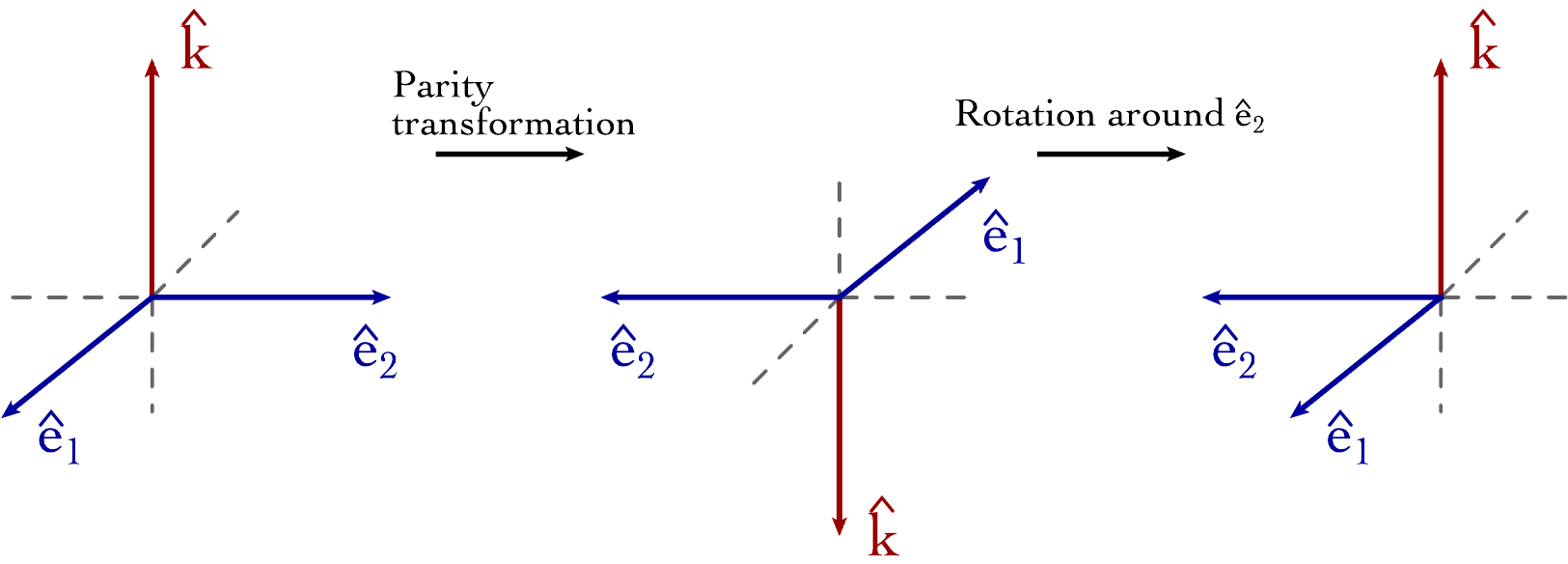}
\end{center}
\vspace{-.25 cm}
\caption{\label{fig:LightPol}\em Parity transformation for the triad 
$(\hat{k}, \hat{e}_1, \hat{e}_2)$. Under a genuine parity transformation 
$(\hat{k}, \hat{e}_1, \hat{e}_2) \overset{P}{\to} (-\hat{k}, -\hat{e}_1, -\hat{e}_2)$ (central panel).
A further $\pi$ rotation around the $\hat{e}_2$ axis (right panel), made possible by isotropy of space,
brings the versors to the final configuration $(\hat{k}, \hat{e}_1, -\hat{e}_2)$.
}
\end{figure}
The situation is illustrated in two steps in fig.~\ref{fig:LightPol}. The wave vector $\hat{k}$ flips sign as a consequence of the Fourier space identification $\vec{\nabla} \to i\vec{k}$. The two polarization vectors also flip sign. This is evident in the Coulomb gauge, in which $\vec{E} = i\omega\vec{A}$. 
The 
vector potential $\vec{A}$ inherits the parity transformation property of the electric field, $\vec{E} \overset{P}{\to} -\vec{E}$. The Coulomb gauge is very useful because it exhibits the physical degrees of freedom:
the 3 components of $\vec{A}$ satisfy the constraint $\vec{\nabla}\cdot \vec{A} = 0$, leaving behind the
2 degrees of freedom that can be identified with the polarization states of
the photon. This means that one can write (for some numbers $a_{i=1,2}$ left unspecified) $\vec{A} = \sum_{i=1,2} a_{i}\hat{e}_i$, and the parity transformation of $\hat{e}_{i=1,2}$ follows from $\vec{A} \overset{P}{\to} -\vec{A}$.  Finally, because of isotropy of space, only the relative orientation between vectors really matters. We can therefore apply a $\pi$ rotation around the $\hat{e}_2$ axis  in order to get the final parity transformation quoted above,
$(\hat{k}, \hat{e}_1, \hat{e}_2) \overset{P}{\to} (\hat{k}, \hat{e}_1, -\hat{e}_2)$. 
This specific choice suggests to use left- and right-handed circular polarization vectors defined by
$\hat{e}_{\rm L,R} \equiv (\hat{e}_1 \mp i\hat{e}_2)/\sqrt{2}$ since under parity 
$\hat{e}_{\rm L,R} \overset{P}{\to} \hat{e}_{\rm R,L}$.
In the absence of parity violation, 
there should be no difference in the physical properties of a right- and a left-handed 
circularly  polarized electromagnetic wave. 
This  discussion is of course a trivial consequence of parity invariance of electromagnetism.

The photon coupling in eq.~(\ref{eq:AxionPhotonL}) does not respect parity, 
since $\vec{E} \overset{P}{\to} -\vec{E}$ and $\vec{B} \overset{P}{\to} \vec{B}$.
This implies that the left and right components of an electromagnetic wave traveling through an axion background  should experience different physical effects.
This is precisely what we shall explore in this paper considering the axion cloud 
surrounding a Kerr black hole as an optically active medium. 

This paper is structured as follows.
In section~\ref{sec:AxionCloud} we discuss general aspects of black hole superradiance with a particular emphasis on the conditions that allow for an analytical approach.
 In section~\ref{sec:Bending} we compute the polarization-dependent bending that 
a ray of light experiences by traveling through an axion cloud, and in section~\ref{sec:Discussion} we discuss the phenomenological relevance of our result. Finally, we conclude in section~\ref{sec:Conclusions}. 
Further technical details are provided in the appendices.

\section{Axion clouds around rotating black holes}\label{sec:AxionCloud}

The massive Klein-Gordon equation
\begin{equation}
\Box\Phi = \mu^2 \Phi
\end{equation}
in a Kerr background
\begin{equation}\label{eq:KerrMetric}
ds_{\rm Kerr}^2 = -\left(
1-\frac{2Mr}{\Sigma} 
\right)dt^2 + \frac{\Sigma}{\Delta}dr^2 - \frac{4raM}{\Sigma}s^2_{\theta} d\phi dt 
+ \Sigma d\theta^2 + \left[
(r^2 + a^2)s^2_{\theta} + \frac{2rMa^2}{\Sigma}s^4_{\theta}
\right]d\phi^2~,
\end{equation}
where $\Sigma = r^2 + a^2c^2_{\theta}$, $\Delta = (r - r_+)(r- r_-)$, $r_{\pm} = M(1\pm \sqrt{1-\tilde{a}^2})$, $a = J/M$, $\tilde{a} = a/M$,
admits the existence of quasi-bound states, as we shall briefly review in the following. 

We use the short-hand notation $s_\theta\equiv \sin\theta$, $c_\theta\equiv \cos\theta$,
and $(t,r,\theta,\phi)$ are the usual Boyer-Lindquist coordinates.
 We work in natural units in which Planck's constant $\hbar$, the
speed of light $c$, and Newton's constant $G_N$
are set to one.  Occasionally, we will reintroduce $G_N$ to make some equations more transparent.

 The massive Klein-Gordon equation in the Kerr
background allows separation of variables\footnote{This property follows from the fact that 
the Kerr metric
admits -- among its mysterious ``miracles''~\cite{Chandra} -- the existence of
 a Killing-Yano tensor~\cite{Carter:1968rr,Walker:1970un}.} with the following simple ansatz for the scalar  field~\cite{Detweiler:1980uk}
\begin{equation}\label{eq:FullKG}
\Phi(t,r,\theta,\phi) = \sum_{l,m}e^{im\phi}S_{lm}(\theta)e^{-i\omega t}R_{nl}(r)~.
\end{equation}
The angular equation  
defines the spheroidal harmonics $S_{lm}(\theta)$~\cite{Berti:2005gp}. 
The angular eigenvalues $\lambda_{lm}$ are approximated by 
\begin{equation}
\lambda_{lm} \simeq l(l+1) + \frac{2c^2\left[
m^2 - l(l+1) + 1/2
\right]}{(2l -1 )(2l + 3)}~,
\end{equation}
where the so-called degree of spheroidicity $c^2$ is defined by $c^2 \equiv a^2(\omega^2 - \mu^2)$. 
The radial part, on the contrary, reduces to a Schr\"odinger-like problem. 
Defining the Regge-Wheeler tortoise coordinate $dr^* = [(r^2 + a^2)/\Delta] dr$, and rescaling the radial function according to $u_{nl}(r^*) = (r^2 + a^2)^{1/2}R_{nl}(r)$, 
the radial equation reads 
\begin{equation}\label{eq:KGTortoise}
\frac{d^2u}{dr^{* 2}} +\left[
\omega^2 - V(\omega)
\right]u = 0~,
\end{equation}
where the potential is
\begin{equation}\label{eq:GWPotential}
V = \frac{\Delta \mu^2}{r^2 + a^2} +
\frac{4Mr am\omega - a^2 m^2 +\Delta\left[
\lambda_{lm} + (\omega^2 - \mu^2)a^2
\right]}{(r^2 + a^2)^2}  +
\frac{\Delta(3r^2 - 4Mr + a^2)}{(a^2 + r^2)^3}
- \frac{3r^2 \Delta^2 }{(r^2 + a^2)^4}~.
\end{equation}
The relation between the tortoise coordinate $r^*$ and the ordinary radial coordinate $r$ is
\begin{equation}
r^* = r + \frac{2Mr_+}{(r_+ - r_-)}\ln\left(
\frac{r}{r_+} - 1
\right) - \frac{2Mr_-}{(r_+ - r_-)}\ln\left(
\frac{r}{r_-} - 1
\right)~.
\end{equation}
The radial equation must be solved with the following 
boundary conditions 
\begin{equation}\label{eq:Boundaries}
R  \underset{ r^* \to -\infty }{\sim} e^{-ik_+r^*}~,~~~~~~R  \underset{ r^* \to \infty }{\sim} \frac{1}{r}\,e^{i(\omega^2 - \mu^2)^{1/2} r^*}~,
\end{equation}
with $k_+ \equiv \omega - m\Omega_H$, being $\Omega_H \equiv a/2Mr_+$ the angular velocity of the Kerr black hole.
Notice that we have purely ingoing waves at the horizon ($r^* = -\infty$ in tortoise coordinate); towards spatial infinity, on the contrary, the solution tends to zero since we are interested in bound states.

The manipulations above 
reduced the problem to the motion of a particle subject to the one-dimensional effective potential in eq.~(\ref{eq:GWPotential}).
\begin{figure}[!htb!]
\centering
\minipage{0.5\textwidth}
  \includegraphics[width=.9\linewidth]{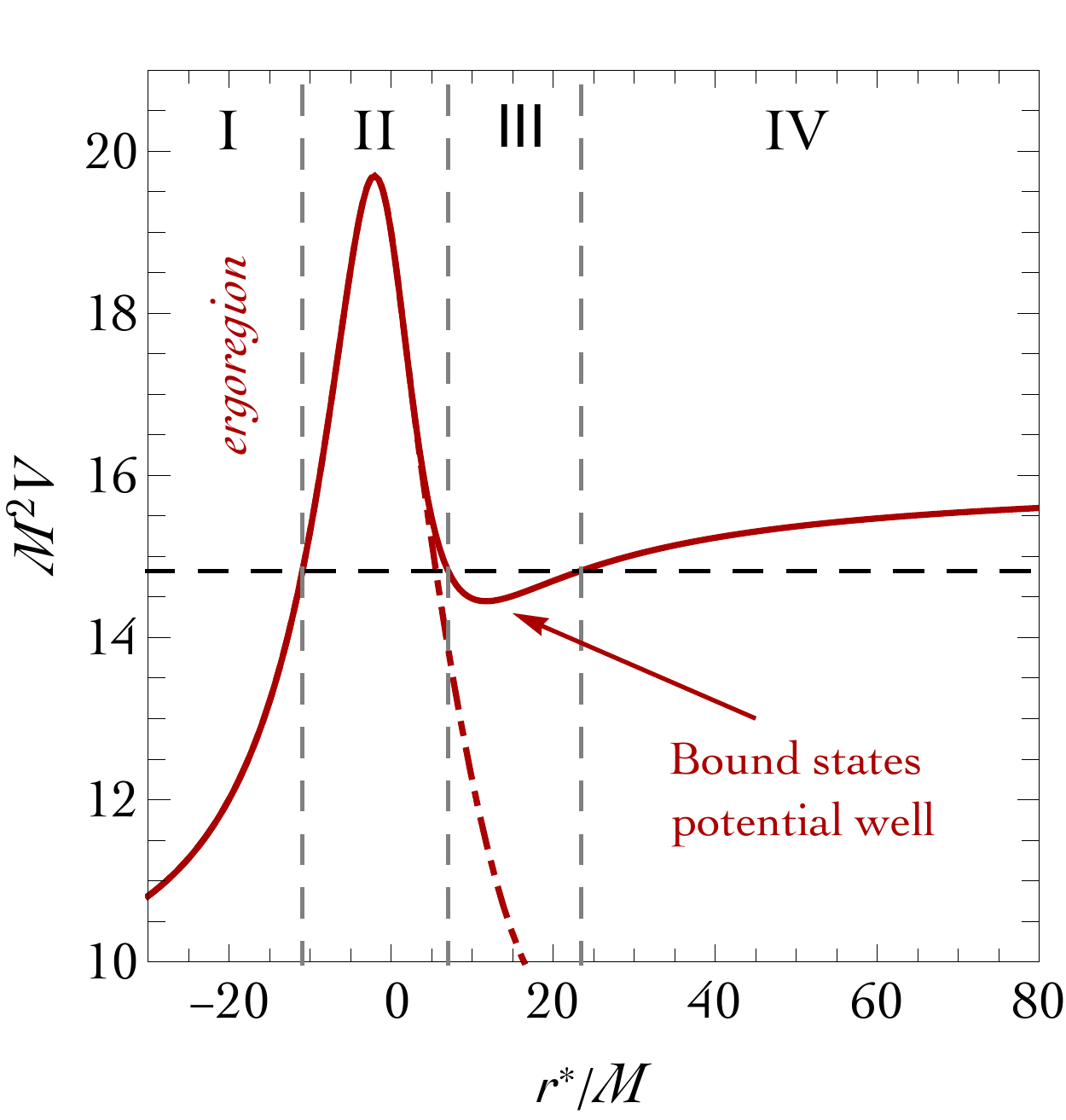}
\endminipage 
\minipage{0.5\textwidth}
  \includegraphics[width=.9\linewidth]{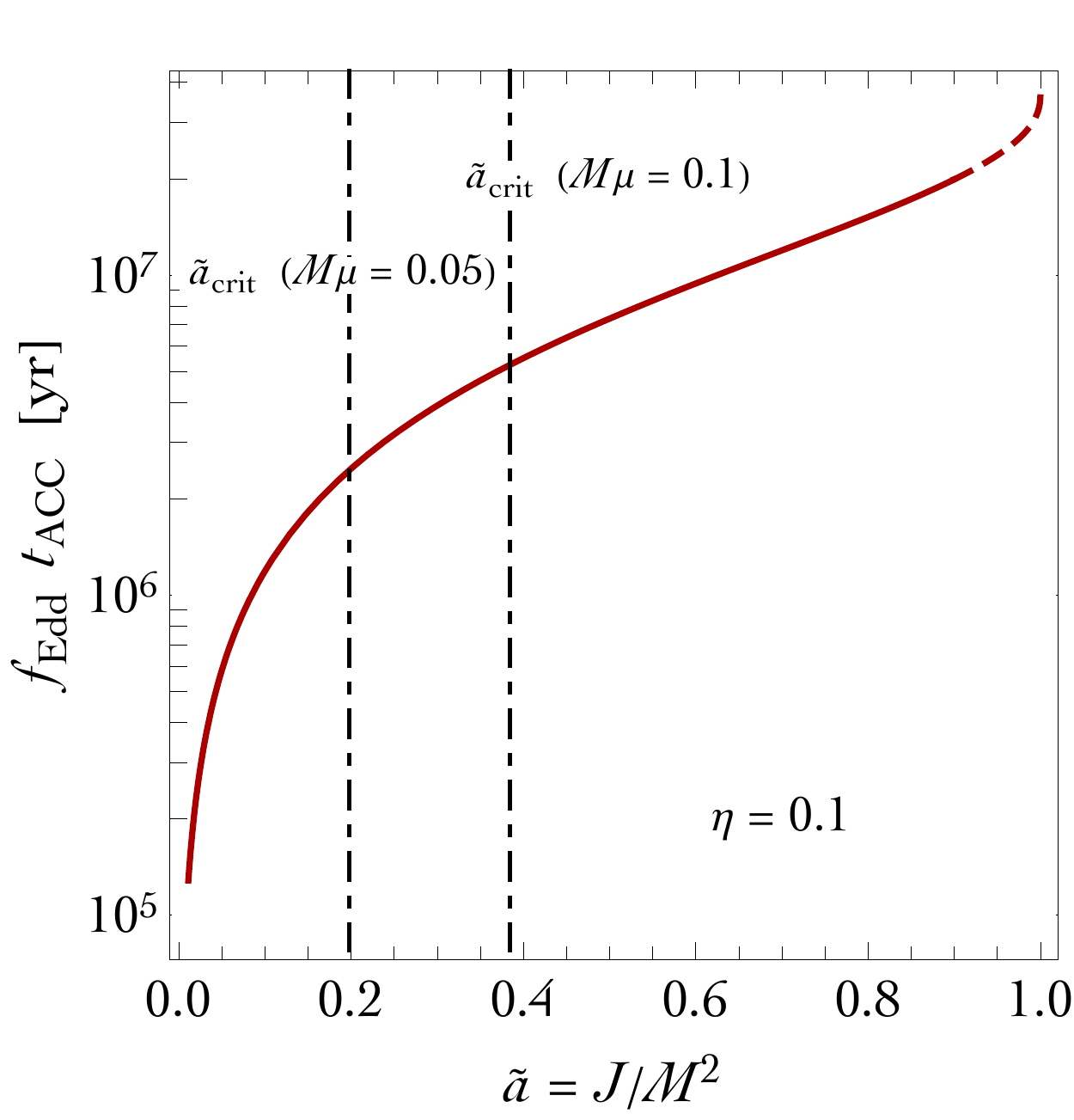}
\endminipage 
\vspace{-.2 cm}
\caption{\label{fig:KGpotential}\em 
Left panel. Effective potential in eq.~(\ref{eq:GWPotential}) as a function of the tortoise coordinate $r^*$.
Right panel. Evolution of the black hole angular momentum due to accretion starting from the Schwarzschild limit. 
Vertical lines mark two critical conditions in eq.~(\ref{eq:CriticalSpin}) for $m=1$ and different values of $M\mu$. The solid red line becomes dashed where the inclusion of radiation is important.
}
\end{figure}
We show the effective potential in the left panel of fig.~\ref{fig:KGpotential}. The presence of the mass term in the Klein-Gordon equation
generates a potential well in region III, allowing for the formation of bound states. Notice that in the massless limit, 
the potential well cannot be formed (dot-dashed red line in the left panel of fig.~\ref{fig:KGpotential}).
Gravitational and centrifugal effects create a potential barrier in region II, and the particle bounded in region III can tunnel 
in the black hole ergoregion, region I. 
If the phase velocity of the purely ingoing wave at the horizon is negative -- that  is if $\omega_R < m \Omega_H$ from the boundary condition in eq.~(\ref{eq:Boundaries}), with $\omega_R \equiv \Re(\omega)$  -- the transmitted wave will carry negative energy into the black hole, and 
the reflected wave will return to infinity with greater amplitude and energy than the incident wave: The superradiance mechanism 
is triggered.

The growth of superradiant instability 
depends on the dimensionless product $M\mu$. This product represents the ratio
between the horizon size of the black hole and the Compton wavelength $\lambda_{\rm Compton}$ of the scalar field
\begin{equation}
M\mu \equiv \frac{G_N M \mu}{\hbar c} \sim \frac{r_+}{\lambda_{\rm Compton}}~.
\end{equation}
Two limits are commonly used, $M\mu \ll 1$ and $M\mu \gg 1$. 
The crucial difference is the growth rate of  bound states. Parametrically, we have the following order-of-magnitude 
estimates~\cite{Detweiler:1980uk,Zouros:1979iw,Arvanitaki:2010sy} 
\begin{equation}\label{eq:FastSuperradiance}
 \tau  \underset{  M\mu \ll 1}{\approx}  \frac{M}{(M\mu)^9}~,~~~~~~~~~~~~~~\tau  \underset{  M\mu \gg 1}{\approx}  10^7 e^{3.7(M\mu)} M~.  
\end{equation}
In the limit $M\mu \ll 1$ the growth of superradiant instability can be as short as $10^2$ s for stellar black holes
\begin{equation}
\tau \sim 10^2\left(\frac{M}{10\,M_{\odot}}\right)
\left(
\frac{0.2}{M\mu}
\right)^9\,{\rm s}~,
\end{equation}
while in the opposite limit the presence of the $e$-folding makes  the instability insignificant 
for astrophysical black holes.

In the following we assume the small $M\mu$ limit, with
\begin{equation}\label{eq:SmallM}
M\mu = 7.5\times 10^{-2} 
\times \left(
\frac{M}{10\,M_{\odot}}
\right) \times \left(
\frac{\mu}{10^{-12}\,{\rm eV}}
\right)~.
\end{equation} 
The small $M\mu$ limit allows for a simple analytical understanding of superradiance.

In the small $M\mu$ limit, the eigenvalue problem for the radial equation 
admits an hydrogenic-like solution $\omega \equiv \omega_R + i\omega_I$~\cite{Detweiler:1980uk}
\begin{eqnarray}
\omega_R &\simeq& \mu - \frac{\mu}{2}\left(
\frac{M\mu}{l + n + 1}
\right)^2~, \label{eq:LambdaApprox1} \\ 
\omega_I &\simeq& \mathcal{F}_{nl} \frac{\left(
M\mu
\right)^{4l + 5}}{M}\left(
\frac{a m}{M} - 2\mu r_+
\right)
\prod_{j = 1}^{l}
\left[
j^2\left(
1-\frac{a^2}{M^2}
\right)+\left(
\frac{am}{M} - 2\mu r_+
\right)^2
\right]~,\label{eq:LambdaApprox2}
\end{eqnarray}
with 
\begin{equation}
\mathcal{F}_{nl} \equiv \frac{2^{4l + 2}(2l + 1 + n)!}{(l+n+1)^{2l+4}n!}
\left[
\frac{l!}{(2l)!(2l+1)!}
\right]^2~.
\end{equation}
The eigenfrequencies are, in general, complex, and the superradiance condition reads 
\begin{equation}\label{eq:CriticalSpin}
a_{\rm crit} \sim \frac{2\mu r_+M}{m}~.
\end{equation}
When $a > a_{\rm crit}$, the imaginary part of $\omega$
becomes positive:  The corresponding modes increase in time, signaling an instability of the
Kerr black hole in the presence of the massive scalar field.

In the small $M\mu$ limit, the radial eigenfunction reads~\cite{Detweiler:1980uk,Yoshino:2013ofa} (see also appendix~\ref{app:Numerics})
\begin{equation}\label{eq:RadialApprox}
R_{nl}(r) = A_{nl}g(\tilde{r})~,~~~~~~
g(\tilde{r}) \equiv \tilde{r}^l e^{-\tilde{r}/2}{\rm L}_{n}^{2l + 1}(\tilde{r})
~,~~~~~~
\tilde{r} \equiv \frac{2rM\mu^2}{l + n + 1}~,
\end{equation}
with ${\rm L}_{n}^{2l + 1}(\tilde{r})$ the Laguerre polynomials. 
In analogy with the hydrogen atom, the combination $\nu \equiv l+n+1$ defines the principal quantum number.

It is important to notice that the size of the cloud can be estimated as~\cite{Arvanitaki:2010sy}
\begin{equation}\label{eq:rCloud}
r_{\rm cloud} \sim \frac{(l + n + 1)^2 M}{(M\mu)^2}\sim 
(l + n + 1)^2\times 1.5\times 10^3\left(
\frac{M}{M_{\odot}}
\right)\left(
\frac{0.1}{M\mu}
\right)^2\,{\rm km}~.
\end{equation}
It implies that the cloud extends way beyond the horizon, where rotation effects can be neglected. 
In this limit the spheroidal harmonics $S_{lm}(\theta)$ reduce to the flat space spherical harmonics.

As clear from the previous discussion, superradiance is a dynamical process.  It is therefore crucial to specify  
what are the assumption underlying our analysis. 
The physical setup we have in mind is the following. 

\begin{figure}[!htb!]
\begin{center}
\minipage{0.5\textwidth}
  \includegraphics[width=.9\linewidth]{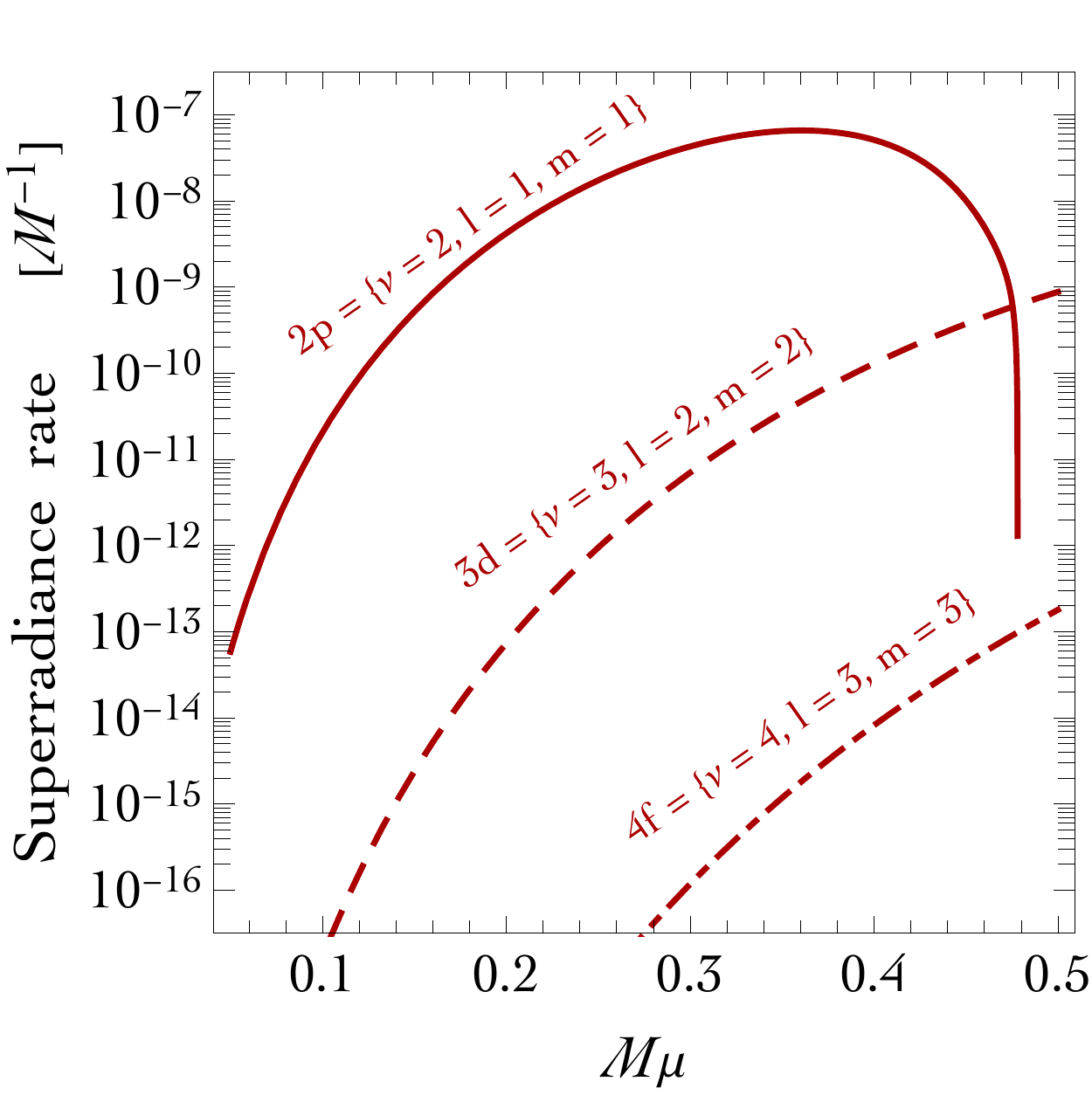}
\endminipage 
\minipage{0.5\textwidth}
  \includegraphics[width=.9\linewidth]{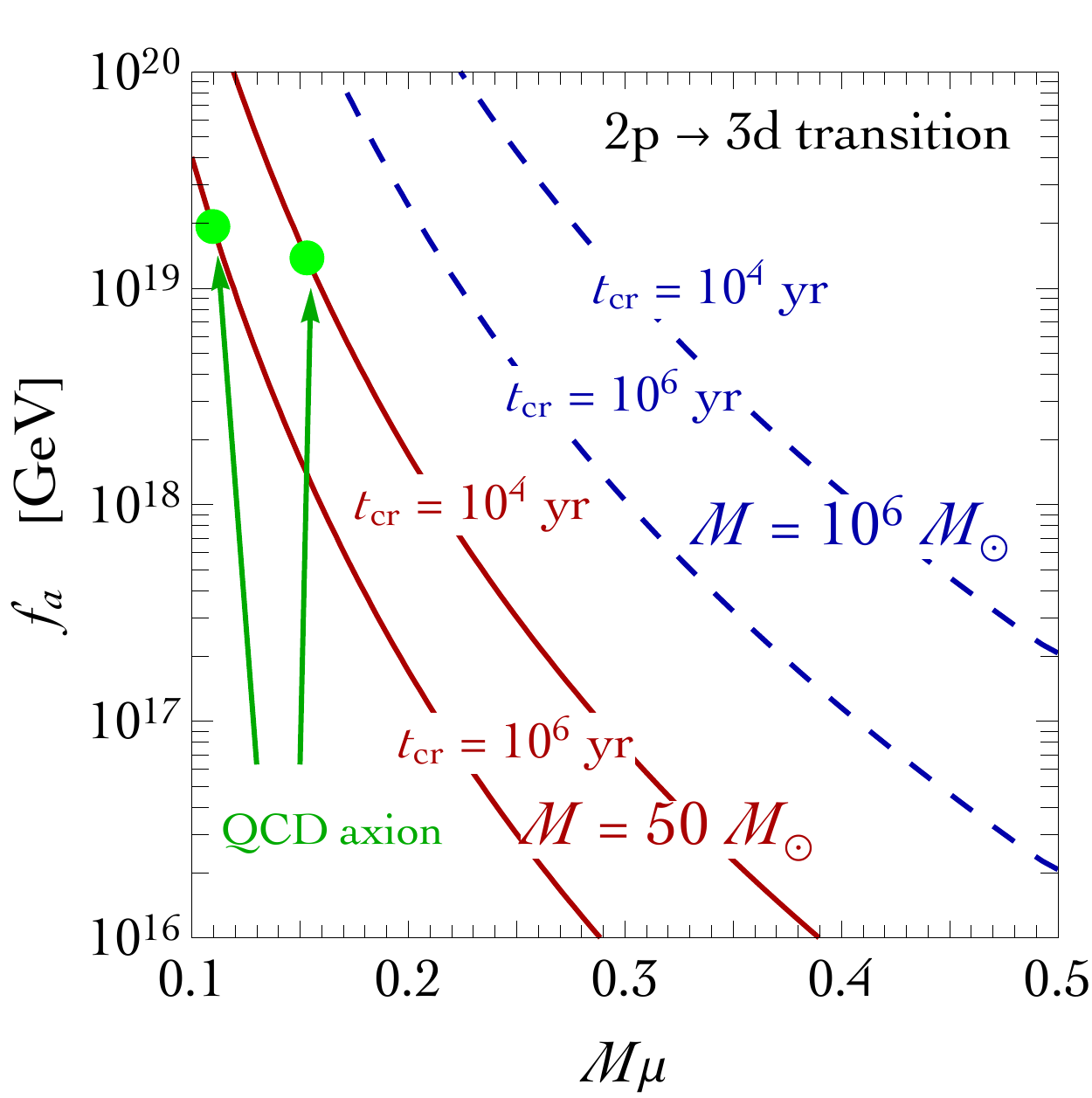}
\endminipage 
\end{center}
\vspace{-.2 cm}
\caption{\label{fig:TimeScale}\em Left panel. Superradiant rates $\omega_I$ as a function of the dimensionless parameter $M\mu$ for different levels.
Right panel. Time required for an axion cloud in the $2p$ level around a Kerr
 black hole with mass $M = 50\,M_{\odot}$ (red solid lines) and 
$M = 10^6\,M_{\odot}$ (blue dashed lines) to 
trigger a superradiant regime in the next $3d$ level
as a function of the axion decay constant $f_a$ and for different values of the parameter $M\mu$.
For each $M\mu$, we compute the critical spin in eq.~(\ref{eq:CriticalSpin}) and the rates
 $\Gamma_{\rm 1s}$ and $\Gamma_{\rm 3d}$ using the  frequencies in eq.~(\ref{eq:LambdaApprox2}).
The time scale of the transition is given by eq.~(\ref{eq:TransitionTimeScale}).
For each of the two analyzed black hole masses, 
the value of the parameter $M\mu$ fixes the axion mass $\mu$. In the case of the QCD axion, the latter is related 
to a specific value of the axion decay constant $f_a$ (see eq.~(\ref{eq:QCD}) below). For a stellar black hole with mass 
$M = 50\,M_{\odot}$, this correspondence is indicated in the plot with the green dots.
}
\end{figure}

\begin{enumerate}

\item 
Let us start considering a rotating black hole. In order to trigger the superradiant instability, 
the black hole must spin above the critical value in eq.~(\ref{eq:CriticalSpin}). 
We can not take this condition for granted, given in particular the lack of
  unambiguous experimental informations about black hole spins at birth. However, it is not difficult to imagine physical processes by means of which a black hole, even starting from a slowly-rotating configuration, increases its mass and spin, eventually fulfilling the superradiant condition. The simplest possibility is provided by accretion. Astrophysical black holes are 
  generally surrounded by an accretion disk of matter in the form of gas and plasma, 
  and the inner edge of this disk is
located in the equatorial plane at the position of the innermost stable circular orbit, $r_{\rm ISCO}$. 
From $r_{\rm ISCO}$, because of the pull of gravitational attraction, particles are sucked into the black hole increasing its mass and angular momentum. We can, therefore, ask the following crude question. 
 What is the typical time scale needed to increase, via accretion, the spin of
 a non-rotating Schwarzschild black hole with initial mass $M_{\rm in}$
  to maximally-rotating values?
 The  accretion of a certain amount of rest mass $\Delta M_0$ results into a change of the black hole mass 
  $M$ and spin $J$ given by $\Delta J = l(z,M)\Delta M_0$ and $\Delta M = e(z)\Delta M_0$~\cite{Page:1974he,Thorne:1974ve},\footnote{In our simplified discussion we do not include the contribution from radiation, i.e. the torque produced by photons emitted from the surface of the accretion disk. As shown in~\cite{Thorne:1974ve}, 
 radiation limits the maximum spin to $\tilde{a} \lesssim 0.998$. The inclusion of radiation is, therefore, important to prevent violation of the cosmic censorship hypothesis but it is not crucial for our argument.} where $z\equiv r_{\rm ISCO}/M$, $e(z)$ is the
energy per unit rest mass
and $l(z,M)$ is the
angular momentum per unit
rest mass
for a particle in the vicinity of the black hole. The explicit
 expressions can be found in~\cite{Bardeen:1970zz}. A simple algebraic manipulation leads to a first-order differential equation that can be solved with the Schwarzschild initial condition $z_{\rm in} = 6$. All in all, we find~\cite{Bardeen:1970zz}
 \begin{equation}\label{eq:Bardeen}
 \tilde{a}(M) = \left(\frac{2}{3}\right)^{1/2}
 \frac{M_{\rm in}}{M}\left\{
 4-\left[
 18\left(
 \frac{M_{\rm in}}{M}
 \right)^2 - 2
 \right]^{1/2}
 \right\}~.
 \end{equation}
To fix ideas, eq.~(\ref{eq:Bardeen}) implies, for instance, that $\tilde{a} = 0.6$ when $M/M_{\rm in} \simeq 1.25$. Having set the relation between mass and spin, we now need an expression for the mass accretion rate. Following~\cite{Brito:2014wla}, we assume the mass accretion rate to be proportional to the Eddington rate
 $\dot{M} =  f_{\rm Edd}\dot{M}_{\rm Edd} = f_{\rm Edd}(4\pi G_N M m_p/\eta \sigma_{T})$, where $\eta$ is the efficiency, $m_p$ the proton mass and $\sigma_{T} \approx 1.7\times 10^3$ GeV$^{-2}$ the Thompson cross-section. We take $\eta = 0.1$. The reader should keep in mind that this is a very conservative estimate. 
 It is indeed possible to imagine values of $\dot{M}$ much greater than the ones inferred by using the Eddington formula  by making the accretion disk physically thick, and with low density.
By integrating the mass accretion formula we find the following expression for the accretion time $t_{\rm ACC}$
\begin{equation}
\ln\frac{M}{M_{\rm in}} = f_{\rm Edd}\left(
\frac{4\pi G_N m_p}{\eta \sigma_T}
\right)t_{\rm ACC}~,
\end{equation}
where in the left-hand side the ratio $M/M_{\rm in}$ can be obtained by inverting eq.~(\ref{eq:Bardeen}).  
In the right panel of fig.~\ref{fig:KGpotential} we show the product $f_{\rm Edd}t_{\rm ACC}$ in years (yr) 
as a function of the black hole spin. 
As mentioned above, the computation of $t_{\rm ACC}$ is subject to some astrophysical uncertainty, 
and the only intent of our plot is to show that, even starting from the borderline case of 
a Schwarzschild black hole, it is possible to reach critical values of spin
in a finite amount of time. 
We refer the reader to~\cite{Brito:2014wla} 
for a more detailed numerical study about the interplay between accretion and superradiance, and for the rest of the paper we will assume that the scalar cloud is not directly coupled to the disk.

\item
When the condition $a > a_{\rm crit}$ is satisfied, the black hole rapidly loses its spin favoring the growth of the axion cloud. The cloud sprouts up from an initial seed that can be simply provided by a quantum fluctuation of the vacuum, as suggested in~\cite{Arvanitaki:2014wva}. 
 En route, we also note that Kerr black hole itself may naturally provide a source term for the axion field. 
This is because the Kerr metric in eq.~(\ref{eq:KerrMetric}) has non-vanishing 
Hirzebruch signature density $R\tilde{R}$~\cite{Frolov:1998wf}. By explicit computation, we find
\begin{equation}\label{eq:QuantumSource}
\frac{1}{2}R\tilde{R} \equiv  
\frac{1}{2}\varepsilon^{\alpha\beta\mu\nu} R_{\rho\lambda\alpha\beta}R_{\mu\nu}^{~~\rho\lambda} =
\frac{\epsilon^{\alpha\beta\mu\nu}}{2\sqrt{-g}}R_{\rho\lambda\alpha\beta}R_{\mu\nu}^{~~\rho\lambda} = 
\frac{288\,\tilde{a} M^3 \cos\theta}{r^7} + \mathcal{O}(\tilde{a}^2)~.
\end{equation}
$R\tilde{R}$ is proportional to the spin, and vanishes for a Schwarzschild black hole. 
If the electromagnetic field is quantized in a gravitational background with such property, 
the pseudo-scalar combination $F_{\mu\nu}\tilde{F}^{\mu\nu}$ acquires 
a non-vanishing expectation value $F_{\mu\nu}\tilde{F}^{\mu\nu} = R\tilde{R}/48 \pi^2$~\cite{Dolgov:1987yp} which, in turn, acts like a background source term for the axion field via the usual electromagnetic coupling.
After this digression, let us now go back to the growth of the axion cloud.
 In the left  panel of fig.~\ref{fig:TimeScale} we show the superradiance rates in eq.~(\ref{eq:LambdaApprox2}) -- in units of $M^{-1}$ --  for different levels. 
 In the small $M\mu$ limit the fastest superradiant level is the $2p$ level with $n = 0$ and $l = m = 1$. 
 The black hole loses its spin by populating the $2p$ shell while all the remaining ones can be neglected.
 As already noticed in eq.~(\ref{eq:FastSuperradiance}), this process can be as short as $10^2$ s for stellar black holes.
 
 \item
The spin-down of the black hole continues until it reaches the threshold value given by eq.~(\ref{eq:CriticalSpin}) with $m = 1$. 
The imaginary part in eq.~(\ref{eq:LambdaApprox2}) vanishes, and the spin-down process terminates.
The black hole remains in this state for a period of time that can be very long. 
Indeed, the next $3d$ level of the axion cloud 
 does not start being populated until a large enough number of axions dissipate from the $2p$ level.
 In this respect, annihilation into gravitons and annihilation into unbounded axions due to self-interactions are the most efficient 
 processes~\cite{Arvanitaki:2010sy}.
 As soon as the the cloud mass drops below a critical value, superradiance becomes operative again, and the black hole rapidly 
 travels to the next level. As discussed in~\cite{Arvanitaki:2010sy}, the time required for an axion cloud in the $2p$ level to dissipate such that 
  the next superradiant level can start being populated  can be extremely long -- specially in the small $M\mu$ limit.
To give a concrete idea, the annihilation time -- considering the  $2p\to 3d$ transition --  can be computed as follows.
 We start writing in full generality the time evolution of the axion population in the $2p$ level due to axion annihilation into gravitons as
 $dN/dt = -\Gamma_{\rm ann}N^2$. The annihilation rate $\Gamma_{\rm ann}$ is given by 
 \begin{equation}
 \Gamma_{\rm ann} = \frac{1}{2\omega N^2}\int d\Omega \frac{dP}{d\Omega}~,
 \end{equation}
 where $N$ is the number of axions and $\int d\Omega\,dP/d\Omega \equiv dE_{\rm GW}/dt$ is the energy per unit of time emitted into gravitational radiation.
 When the superradiance condition is satisfied the imaginary part of $\omega$ vanishes, and in the small 
 $M\mu$ limit we have $\omega_{\rm R}\approx \mu$. The computation of $dE_{\rm GW}/dt$ cannot be performed in flat space because the leading term in the small $M\mu$ expansion accidentally cancels. We therefore use the corresponding expression derived in the Schwarzschild background metric~\cite{Brito:2014wla} 
 \begin{equation}\label{eq:RadiationRate}
\frac{dE_{\rm GW}}{dt} =
\frac{484 + 9\pi^2}{23040}
\left(
\frac{M_S^2}{M^2}
\right)(M\mu)^{14}~,
 \end{equation}
 where $M_S$ is the mass of the axion cloud. Furthermore, 
 since axions are non-relativistic, we can write $M_S = N\mu$. 
  Eq.~(\ref{eq:RadiationRate}) is in good agreement with the 
  computation recently performed
  in~\cite{Brito:2017wnc,Brito:2017zvb} using the Teukolsky formalism in the
fully relativistic regime. We can now integrate $dN/dt = -\Gamma_{\rm ann}N^2$, and find
\begin{equation}\label{eq:CriticalTime}
N(t) = \frac{N(0)}{1 + \Gamma_{\rm ann}N(0)t}\approx \frac{1}{\Gamma_{\rm ann}t}~.
\end{equation}
In order to proceed further, we use the condition according to which 
the $3d$ level starts being populated when the number of axions in the $2p$ level drops 
below the value~\cite{Arvanitaki:2010sy} 
\begin{equation}\label{eq:Ncondition}
N \lesssim \frac{16\pi f_a^2 M^2}{(M\mu)^3}\left|
\frac{\Gamma_{\rm 3d}}{\Gamma_{\rm 1s}}
\right|^{1/2}~.
\end{equation}
The presence in eq.~(\ref{eq:Ncondition}) of the damping rate related to the level $1s$ is due 
to the effect of axion non-linearities. These interactions are responsible for level mixing, and 
introduce a superposition of the $2p$ level with the non-superradiant $1s$ mode.
In our example -- remember that we are considering a black hole spin such that the superradiant
condition in eq.~(\ref{eq:CriticalSpin}) vanishes for the $2p$ level -- the frequency of the $1s$ mode has a negative imaginary part, and 
the level is damped. In the small $M\mu$ limit 
we compute the rate $\Gamma_{i}$ using the imaginary frequencies in eq.~(\ref{eq:LambdaApprox2}).
The condition derived in eq.~(\ref{eq:Ncondition}) defines, plugged into eq.~(\ref{eq:RadiationRate}), the critical time scale
\begin{equation}\label{eq:TransitionTimeScale}
t_{\rm cr}\simeq \left(
\frac{720}{484 + 9\pi^2}
\right)\frac{M}{\pi f_a^2 (M\mu)^{12}}\left|
\frac{\Gamma_{\rm 1s}}{\Gamma_{\rm 3d}}
\right|^{1/2}~.
\end{equation}
In fig.~\ref{fig:TimeScale} we show the time in years to depopulate the level $2p$ for two representative value of black hole
 mass, $M = 50\,M_{\odot}$ and $M = 10^6\,M_{\odot}$, as a function of the axion coupling $f_a$ and the parameter $M\mu$.
 From this estimate it is clear that in the small $M\mu$ limit the axion cloud 
 can remain stuck for a very long time in the $2p$ level.
 It is therefore reasonable to focus on the values $l = m = 1$, $n = 0$. Motivated by these arguments,
 we adopt this assumption in the rest of the paper.
   
\end{enumerate}

There are two scales in the problem, the oscillation time $\tau_S = 1/\omega_R$ and the instability growth time scale
$\tau \equiv 1/\omega_I$. In the small $M\mu$ limit we have
\begin{eqnarray}
\omega_R &=& \mu - \frac{\mu}{2}\left(
\frac{M\mu}{2}
\right)^2 \approx \mu~,\label{eq:RealPart}\\
\omega_I &=&   \frac{1}{48 M}\left(
\frac{a}{M} - 2\mu r_+
\right)\left(
M\mu
\right)^{9}\approx \frac{\left(
M\mu
\right)^{9}}{M}~.
\end{eqnarray}
As a consequence
\begin{equation}
\frac{\tau}{\tau_S} \approx \frac{1}{(M\mu)^8} \gg 1~~~~\Longrightarrow~~~~\tau \gg \tau_S~.
\end{equation}
We can therefore assume a stationary cloud, and write 
\begin{equation}\label{eq:Cloud}
\Phi(t,r,\theta,\phi) = A_0 g(\tilde{r}) \cos\left(
\phi - \omega_R t
\right)\sin\theta~,~~~~~A_{0} \equiv A_{01}~.
\end{equation}
Notice that we focused on a real scalar cloud, since we have in mind the axions.
The amplitude $A_0$ can be expressed in terms of the mass $M_S$ of the scalar cloud~\cite{Brito:2014wla}. 
In full generality, we write 
\begin{equation}\label{eq:MS}
M_S = \int \rho\,r^2 dr \sin\theta d\theta d\phi~,
\end{equation}
with $\rho = -T_0^0$. The energy density $\rho$ can be directly computed from the definition of the stress-energy tensor 
\begin{equation}
T^{\mu\nu}(\Phi) = (D^{\mu}\Phi)(D^{\nu}\Phi) - g^{\mu\nu}\left[
\frac{g^{\rho\sigma}}{2}(D_{\rho}\Phi)(D_{\sigma}\Phi) + V(\Phi)
\right]~,
\end{equation}
where $V(\Phi) = \mu^2 \Phi^2/2$. Assuming flat space -- see comment below eq.~(\ref{eq:rCloud}) -- we find 
\begin{eqnarray}
\rho &=& \frac{A_0^2}{2r^2}\left\{
\mu^4M^2r^2g^{\prime}(\tilde{r})^2\sin^2\theta\cos^2(\phi - \omega_R t)
\right. \nonumber \\
&+&\left.
g(\tilde{r})^2\left[
\cos^2(\phi - \omega_R t)\left(
\cos^2\theta + \mu^2 r^2\sin^2\theta
\right) + \sin^2(\phi - \omega_R t)\left(
1 + \omega_R^2 r^2 \sin^2\theta
\right)
\right]
\right\}~.
\end{eqnarray}
The integral in eq.~(\ref{eq:MS}) can be straightforwardly computed, and we find
\begin{equation}
M_S = \frac{2\pi A_0^2}{3M\mu^2}\left[
2\mathcal{I}_0 + \mathcal{I}_2^{\prime} + \frac{2\mathcal{I}_2}{M^2 \mu^2}
\right]~,~~~~~\mathcal{I}_n = \int_0^{\infty}dx\,x^n g(x)^2~,~~~~~\mathcal{I}_n^{\prime} = 
\int_0^{\infty}dx\,x^n g^{\prime}(x)^2~.
\end{equation}
In the small $M\mu$ limit we have
\begin{equation}
A_0^2 = \frac{3}{4\pi \mathcal{I}_2}\left(
\frac{M_S}{M}
\right)(M\mu)^4~,~~~~{\rm with}~~~\mathcal{I}_2 = 24~.
\end{equation}
The scalar cloud in eq.~(\ref{eq:Cloud}) becomes
\begin{equation}\label{eq:CloudFinal}
\Phi = \sqrt{
\frac{3}{4\pi \mathcal{I}_2}\left(
\frac{M_S}{M}
\right)
}(M\mu)^2 g(\tilde{r})\cos(\phi - \omega_R t)\sin\theta~.
\end{equation}
Considering for definiteness the value of the cloud at $r_{\rm cloud}$ in eq.~(\ref{eq:rCloud}), we have 
\begin{equation}
r_{\rm cloud} \sim \frac{4M}{(M\mu)^2}~,~~~\tilde{r}_{\rm cloud} \sim 4~~~~~\Longrightarrow~~~~~g(\tilde{r}_{\rm cloud}) \sim 0.5~.
\end{equation}
Furthermore, notice that the function $g(\tilde{r})$ has a maximum (for $l = 1$) at $\tilde{r}_{\rm max} = 2$.

Before proceeding, let us comment about possible limiting  factors  for  the  size  of  the  cloud,
in particular  the  so-called ``bosenova'' collapse~\cite{Arvanitaki:2010sy}. 
The physics of the bosenova collapse can be summarized as follows.
In the first stage, the energy of the cloud grows by superradiant instability.
As the ratio $M_S/M$ increases,  the field amplitude in eq.~(\ref{eq:CloudFinal}) becomes larger -- eventually saturating the condition $\Phi/f_a \sim 1$.
At this point, the nonlinear self-interaction of the axion field becomes important, and causes a rapid collapse of the cloud.
 The analysis in~\cite{Arvanitaki:2010sy} (see also~\cite{Yoshino:2012kn} for a numerical analysis)  implies the condition
 \begin{equation}\label{eq:Bosenova}
\frac{M_S}{M} \lesssim \frac{2l^4 f_a^2}{(M\mu)^4 M_{\rm Pl}^2}~.
\end{equation} 
In 
the situation where $M\mu$ is small and $f_a$ is large, the right-hand side of eq.~(\ref{eq:Bosenova})
 becomes large. In this case, the axion cloud spins down the black hole 
 to reach the marginal superradiant condition, $\mu = m\tilde{a}/2r_+$, well before the nonlinear self-interaction becomes important. In this case, therefore, 
 the main limiting factor is the initial rotation energy of the black hole.

Finally, we note that the typical axionic hair configurations generated by quantum effects~\cite{Bowick:1988xh,Lee:1991jw,Campbell:1991rz,Campbell:1991kz,Reuter:1991cb} are usually suppressed, if compared with 
eq.~(\ref{eq:CloudFinal}), by the factor 
\begin{equation}
\left(
\frac{M_{\rm Pl}}{M}
\right)^2 \sim \frac{10^{-76}}{(M/M_{\odot})^2}~.
\end{equation}
However, these quantum effects may act as a seed for the axion cloud (see discussion related to eq.~(\ref{eq:QuantumSource})).

\section{Polarization-dependent bending of light}\label{sec:Bending}

The Maxwell field equations in the presence of a background axion field are
\begin{eqnarray} 
\vec{\nabla}\cdot \vec{E} &=& - g_{a\gamma\gamma}\vec{\nabla}\Phi \cdot \vec{B}~,\label{eq:AxionQED1}\\
\vec{\nabla} \times \vec{E}  + \frac{\partial\vec{B}}{\partial t}&=& 0~,\\
\vec{\nabla} \times \vec{B} - \frac{\partial\vec{E}}{\partial t} &=& g_{a\gamma\gamma}\left(
-\vec{E} \times \vec{\nabla}\Phi - \vec{B}\frac{\partial\Phi}{\partial t}
\right)~,\\
\vec{\nabla}\cdot \vec{B} &=& 0~,\label{eq:AxionQED4}
\end{eqnarray}
where $g_{a\gamma\gamma}$ is the effective coupling defined by the Lagrangian density 
\begin{equation}
\mathcal{L}_{a\gamma\gamma} = \frac{g_{a\gamma\gamma}}{4}\Phi F_{\mu\nu}\tilde{F}^{\mu\nu} = - \frac{g_{a\gamma\gamma}}{2}\left(
\partial_{\mu}\Phi
\right)A_{\nu}\tilde{F}^{\mu\nu}~,
\end{equation}
with $\tilde{F}^{\mu\nu} = \epsilon^{\mu\nu\rho\sigma}F_{\rho\sigma}/2$.
The effective coupling $g_{a\gamma\gamma}$ can be related the the axion decay constant $f_a$~\cite{diCortona:2015ldu} 
\begin{equation}\label{eq:AxionDecayConstant}
g_{a\gamma\gamma} = \frac{\alpha_{\rm em}}{2\pi f_a}\left[
\frac{E}{N}
-\frac{2}{3}\left(\frac{4m_d + m_u}{m_d + m_u}\right)\right] = 
 \frac{\alpha_{\rm em}}{2\pi f_a}\left(
\frac{E}{N}
-1.92\right) 
~,
\end{equation}
where
$E/N$ is the model-dependent ratio of the electromagnetic and color anomaly while the second term is a model-independent contribution coming from the minimal coupling to QCD at the non-perturbative level.
The typical axion window is defined by the interval $0.07 \leqslant |E/N - 1.92| \leqslant 7$~\cite{Patrignani:2016xqp}. 
Of particular interest are the reference values $E/N = 8/3$ (as in DFSZ models~\cite{Zhitnitsky:1980tq,Dine:1981rt} or
KSVZ~\cite{Kim:1979if,Shifman:1979if} models with heavy fermions in complete $SU(5)$ representations) and $E/N = 0$ (as in 
KSVZ models with electrically neutral heavy fermions). 
Recently~\cite{DiLuzio:2016sbl}, the aforementioned axion window was  redefined in light of 
precise phenomenological requirements -- such as the 
absence of Landau poles up to the Planck scale or the need to avoid overclosure of the Universe -- related to the representations of the new heavy quarks that are needed in KSVZ-type models to induce the anomalous coupling of the axion with ordinary quarks.
As a result, the window $0.25 \leqslant |E/N - 1.92| \leqslant 12.25$ was singled out in the case of one single pair of new heavy fermions. Furthermore, with the inclusion of additional pairs of new heavy quarks values as large as $E/N = 170/3$ become accessible. 
Note that it is also possible to construct models with multiple scalars in which the value of $g_{a\gamma\gamma}$ 
in eq.~(\ref{eq:AxionDecayConstant}) 
can be made arbitrarily large. We shall further explore this possibility in section~\ref{sec:PhotoPhilic}.

For the QCD axion, the axion mass and decay constant are related by~\cite{diCortona:2015ldu}
\begin{equation}\label{eq:QCD}
\frac{10^{16}\,{\rm GeV}}{f_a} = \frac{\mu}{5.7\times 10^{-10}\,{\rm eV}}~.
\end{equation}
Only space-time gradients of the axion field configuration alter the 
 Maxwell equations, since for a constant axion field $\Phi F_{\mu\nu}\tilde{F}^{\mu\nu}$ becomes a perfect derivative and does not affect the equation of motion. 
We assume that the length scale over which $\Phi$ changes appreciably is much larger than the wavelength $\lambda$ of
the electromagnetic wave. Within this approximation 
we can neglect in eqs.~(\ref{eq:AxionQED1}-\ref{eq:AxionQED4}) terms containing second derivative (or first derivative squared) of $\Phi$~\cite{Harari:1992ea}.
Let us briefly discuss the validity of this assumption.
Considering the radial direction, the characteristic length scale of the cloud can be estimated 
using eq.~(\ref{eq:rCloud}). The condition on the wavelength $\lambda$ reads
\begin{equation}
\lambda \ll r_{\rm cloud} \sim 2.6\times 10^6 \left(
\frac{10\,M_{\odot}}{M}
\right)\left(
\frac{10^{-12}\,{\rm eV}}{\mu}
\right)^2\,{\rm m}~.
\end{equation}
From eq.~(\ref{eq:CloudFinal}), the characteristic length scale of time variation is $\tau_S = 1/\omega_R$; since we are interested in the 
small $M\mu$ limit in which $\omega_R \simeq \mu$, we have the following condition on the wavelength $\lambda$
\begin{equation}
\lambda \ll  \lambda_{\rm Compton} \sim 2\times 10^5\left(
\frac{10^{-12}\,{\rm eV}}{\mu}
\right)\,{\rm m}~.
\end{equation}
Clearly, the conditions $\lambda \ll r_{\rm cloud},\,\lambda_{\rm Compton}$ are verified for wavelength $\lambda$ of astrophysical interest.
The field equations take the form~\cite{Harari:1992ea}
\begin{eqnarray}
\Box\left(
\vec{E} - \frac{g_{a\gamma\gamma}}{2}\Phi\vec{B}
\right) &=& -  \frac{g_{a\gamma\gamma}}{2}\Phi\Box\vec{B}~,\\
\Box\left(
\vec{B} + \frac{g_{a\gamma\gamma}}{2}\Phi\vec{E}
\right) &=&  \frac{g_{a\gamma\gamma}}{2}\Phi\Box\vec{E}~,
\end{eqnarray}
and reduce to the usual electromagnetic wave equations in the limit $g_{a\gamma\gamma} \to 0$.
Photon propagation is described by the following dispersion relation~\cite{Carroll:1989vb}
\begin{equation}\label{eq:Dispersion}
k^4 + g_{a\gamma\gamma}^2(\partial_{\mu}\Phi)(\partial^{\mu}\Phi)k^2 = g_{a\gamma\gamma}^2\left[
k_{\mu}(\partial^{\mu}\Phi)
\right]^2~,
\end{equation}
where $k^{\alpha} = (E_{\gamma}, \vec{k})$ is the four-momentum of the propagating photon. 
We give a derivation of eq.~(\ref{eq:Dispersion}) in appendix~\ref{app:Dispersion}.
At the first order, we have
\begin{equation}\label{eq:DispApp}
E_{\gamma}^2 - |\vec{k}|^2 \approx \pm g_{a\gamma\gamma}
\left[
E_{\gamma} \frac{\partial\Phi}{\partial t}
-
\vec{k}\cdot \vec{\nabla} \Phi
\right]~,
\end{equation}
where the sign $\pm$ corresponds to right- and left-handed circularly polarized waves.
In eq.~(\ref{eq:DispApp}) we used a flat background metric. The gradient of the scalar field, in spherical coordinates, is
\begin{equation}
\vec{\nabla} \Phi = \left(
\frac{\partial \Phi}{\partial r}, \frac{1}{r}\frac{\partial \Ph}{\partial \theta}, \frac{1}{r\sin\theta}\frac{\partial\Phi}{\partial\phi}
\right) \stackrel{\theta = \pi/2}{\longrightarrow}
 \left(
\frac{\partial \Phi}{\partial r}, 0, \frac{1}{r}\frac{\partial\Phi}{\partial\phi}
\right)~,
\end{equation}
where in the last passage we restrict the analysis to the equatorial plane.
Eq.~(\ref{eq:DispApp}) reads 
\begin{equation}\label{eq:FlatBanding}
\left(
\frac{dr}{d\xi}
\right)^2 = E_{\gamma}^2 - \frac{L^2}{r^2} \mp g_{a\gamma\gamma}\left\{
E_{\gamma}\frac{\partial \Phi}{\partial t} - \left[
\left(\frac{dr}{d\xi}\right) \frac{\partial \Phi}{\partial r}  + \frac{L}{r^2}\frac{\partial \Phi}{\partial \phi}
\right]
\right\}~,
\end{equation}
where $\xi$ is the affine parameter while $E_{\gamma}$ and $L$ are, respectively, the conserved energy and angular momentum of the photon, with $k^r \equiv dr/d\xi$, 
$k^{\theta} \equiv d\theta/d\xi = 0$, $k^{\phi} \equiv d\phi/d\xi = L/r^2$.
From eq.~(\ref{eq:CloudFinal}), we have
\begin{eqnarray}
\frac{\partial \Phi}{\partial t}  &=& \sqrt{
\frac{3}{4\pi \mathcal{I}_2}\left(
\frac{M_S}{M}
\right)
}(M\mu)^2 g(\tilde{r})\omega_R \sin(\phi - \omega_R t)~,\\
\frac{\partial \Phi}{\partial r}  &=& \sqrt{
\frac{3}{4\pi \mathcal{I}_2}\left(
\frac{M_S}{M}
\right)
}(M\mu)^2g^{\prime}(\tilde{r})(M\mu)^2\cos(\phi - \omega_R t)~,\\
\frac{1}{r}\frac{\partial \Phi}{\partial \phi}  &=& -\frac{1}{r}\sqrt{
\frac{3}{4\pi \mathcal{I}_2}\left(
\frac{M_S}{M}
\right)
}(M\mu)^2g(\tilde{r})\sin(\phi - \omega_R t)~.
\end{eqnarray}
Notice that natural units can be recovered with the formal substitution $M\to G_N M$.
Considering  the radial distance at $\tilde{r}_{\rm max}$, we have 
\begin{equation}\label{eq:approximation}
\frac{\partial \Phi}{\partial r}\,,\frac{1}{r}\frac{\partial \Phi}{\partial \phi} \sim (M\mu)\frac{\partial \Phi}{\partial t}~~~~~
\stackrel{M\mu \ll 1}{\Longrightarrow}~~~~~
\frac{\partial \Phi}{\partial t}  \gg \frac{\partial \Phi}{\partial r}\,,\frac{1}{r}\frac{\partial \Phi}{\partial \phi}~.
\end{equation}
This relation simplifies the equation for the photon orbit in the presence of the axion background field.
The differential equation for the photon orbit (see appendix~\ref{app:C}) is 
\begin{equation}
\frac{d\phi}{dx} = - \frac{1}{x^2\sqrt{\frac{1}{x_{\rm max}^2}  - \frac{1}{x^2}}} 
\mp
\frac{a(E_{\gamma},x,\phi) - a(E_{\gamma},x_{\rm max},\frac{\pi}{2})}
{2x^2 x_{\rm max}^2\left(\frac{1}{x_{\rm max}^2} - \frac{1}{x^2}\right)^{3/2}}
~,~~~~~~{\rm with}~~a(E_{\gamma},r,\phi) \equiv \frac{g_{a\gamma\gamma}}{E_{\gamma}}
\left.
\frac{\partial\Phi}{\partial t}\right|_{r, \phi}~,
\end{equation}
with dimensionless variable $x \equiv r/M$ (which of course corresponds to $x \equiv r/G_NM$ in natural units), 
and must be integrated between $x_0 = \infty$ and $x_{\rm max} = 2/(M\mu)^2$. 
The choice $x_0 = \infty$ practically means that we are considering a source and an observer at distance much larger than the impact parameter (see appendix~\ref{app:C} for a detailed discussion).

The outcome of this computation is the angular separation 
$|\Delta\phi_+ - \Delta\phi_-|$ between left- and right-handed circularly polarized waves that a ray of light  
experiences by traveling through the axion cloud.

In the following we shall solve this equation for the QCD axion and for a generic ALP.
 In section~\ref{sec:Discussion} we shall explain in more detail what is the phenomenological relevance of 
 our computation.

\subsection{The QCD axion}\label{sec:QCDaxion}

Stellar  black  hole superradiance in the presence of an ultra-light scalar field 
may produce in the next few years spectacular signatures --
both direct and indirect -- in gravitational wave detectors such as Advanced LIGO.  Indirect signatures
refer to the observation of gaps in the spin-mass distribution of final state black holes produced by binary
black hole mergers.  Direct signatures refer to monochromatic gravitational wave signals produced during the
dissipation of the scalar condensate after the superradiant condition is saturated.
In~\cite{Arvanitaki:2014wva} it was shown that spin 
and mass measurements of stellar-size black holes exclude the QCD axion mass window
$6\times 10^{-13} \lesssim \mu\,[{\rm eV}] \lesssim  2\times 10^{-11}$,
  corresponding to $3\times 10^{17}\lesssim f_a\,[{\rm GeV}] \lesssim 10^{19}$.
  It is worth emphasizing that this bound 
  is most likely only indicative since it is based on black hole spin measurements that are
extracted indirectly from X-ray observations of accretion disks in X-ray binaries.  
We only have very few of
such measurements at our disposal, and it is difficult to extract a bound with robust statistical confidence.

As far as direct signatures are concerned, a careful assess of the detection prospects in Advanced LIGO and
LISA was recently proposed in~\cite{Brito:2017wnc,Brito:2017zvb}. 
The outcome of the analysis is that, considering optimistic astrophysical
models for black hole populations,  the gravitational wave signal produced by superradiant clouds of scalar
bosons with mass in the range
\begin{equation}\label{eq:GoldenRange}
2\times 10^{-13} \lesssim \mu~[{\rm eV}] \lesssim   10^{-12}~,
\end{equation}
 is observable -- i.e.  it is characterized by a signal-to-noise 
 ratio larger than the experimental threshold -- by Advanced LIGO. 
 For the QCD axion the mass range in eq.~(\ref{eq:GoldenRange}) 
corresponds to $5.7\times 10^{18} \lesssim f_a\,[{\rm GeV}] \lesssim 2.8\times 10^{19}$.
In the following, we shall adopt the mass interval in eq.~(\ref{eq:GoldenRange}) as benchmark for our analysis in the case of the QCD axion.
However, before proceeding, an important comment is in order.
 For large values of $f_a$ non-perturbative gravitational instantons  become important, as discussed in~\cite{Alonso:2017avz}. If computed in the context of General Relativity, these 
 effects generate a gravitational correction to the axion mass that increases with $f_a$ and, if $f_a \gtrsim 10^{16}$ GeV, overcomes the QCD term in eq.~(\ref{eq:QCD}). This effectively produces a lower limit on the QCD axion mass, $\mu \gtrsim 4.8\times 10^{-10}$ eV~\cite{Alonso:2017avz}. From this perspective, 
 the mass range in eq.~(\ref{eq:GoldenRange}) is theoretically disfavored. 
  As discussed in~\cite{Alonso:2017avz} (see also~\cite{Kallosh:1995hi} for the original formulation of the argument), the computation of non-perturbative gravitational effects -- and as a consequence the validity of the lower limit on $\mu$ -- can be invalidated 
  if the UV completion of General Relativity is weakly coupled since in this case we expect new degrees of freedom to become dynamical even below $M_{\rm Pl}$. For this reason, it is important to keep
   investigating Planckian values of $f_a$  since they may open an indirect window on quantum gravity effects.  

The QCD axion with mass in the range given by eq.~(\ref{eq:GoldenRange}) falls into the so-called ``anthropic'' window. The Peccei-Quinn symmetry is broken before the end of inflation, and the possibility to reproduce the observed dark matter
relic density $\Omega_{\rm DM} h^2 \simeq 0.1$ 
 relies on a fine-tuned choice of the initial misalignment angle $\theta_{\rm in}$. 
  We find $1.19 \lesssim \theta_{\rm in}\times 10^5~[{\rm rad}] \lesssim 3.98$ for the mass interval in eq.~(\ref{eq:GoldenRange}).

\begin{figure}[!htb!]
  \begin{center}
\fbox{The QCD axion}\vspace{-0.1cm}
\end{center}
\minipage{0.55\textwidth}
\centering
  \includegraphics[width=.94\linewidth]{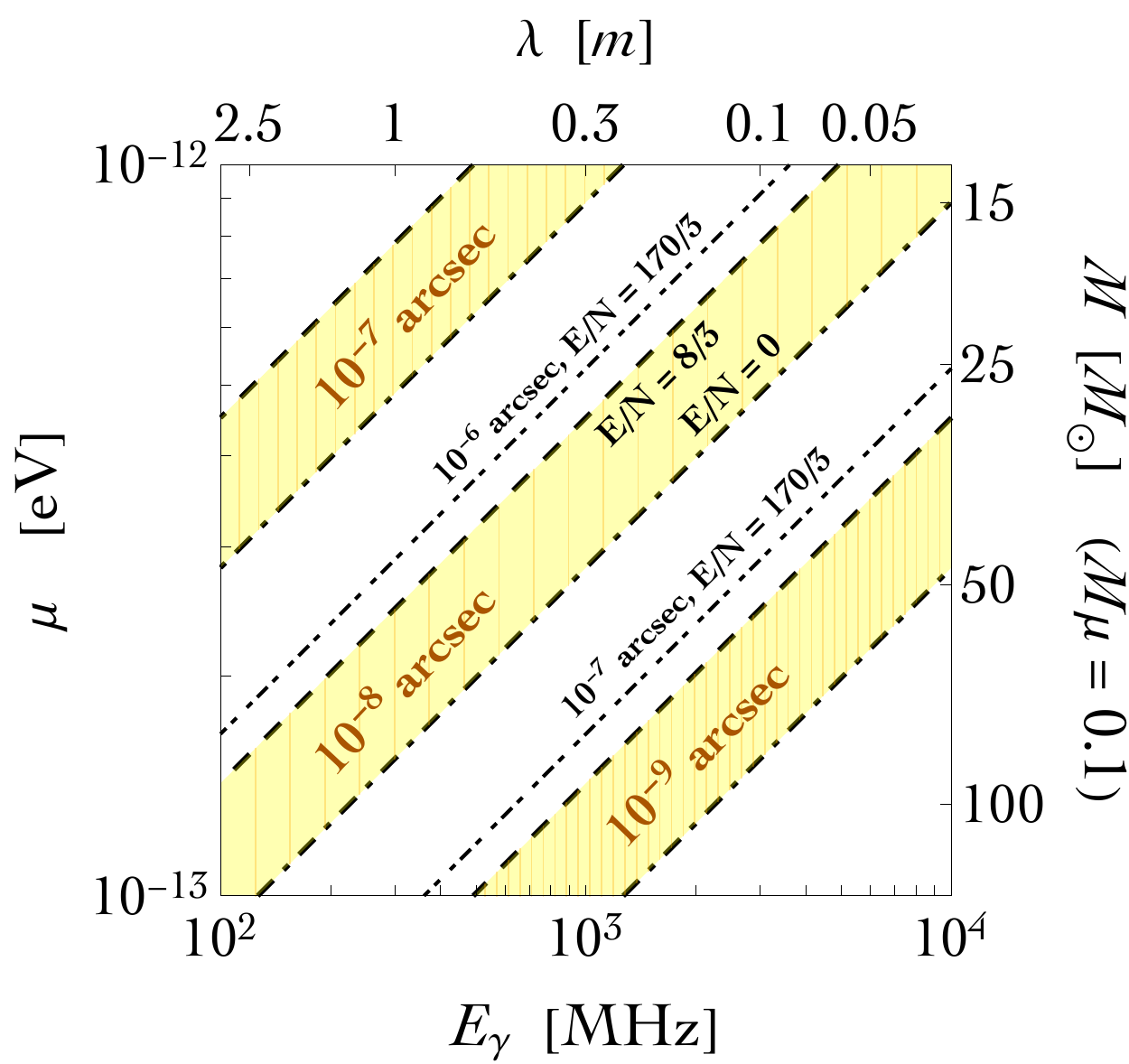}
\endminipage 
\minipage{0.5\textwidth}
  \includegraphics[width=.93\linewidth]{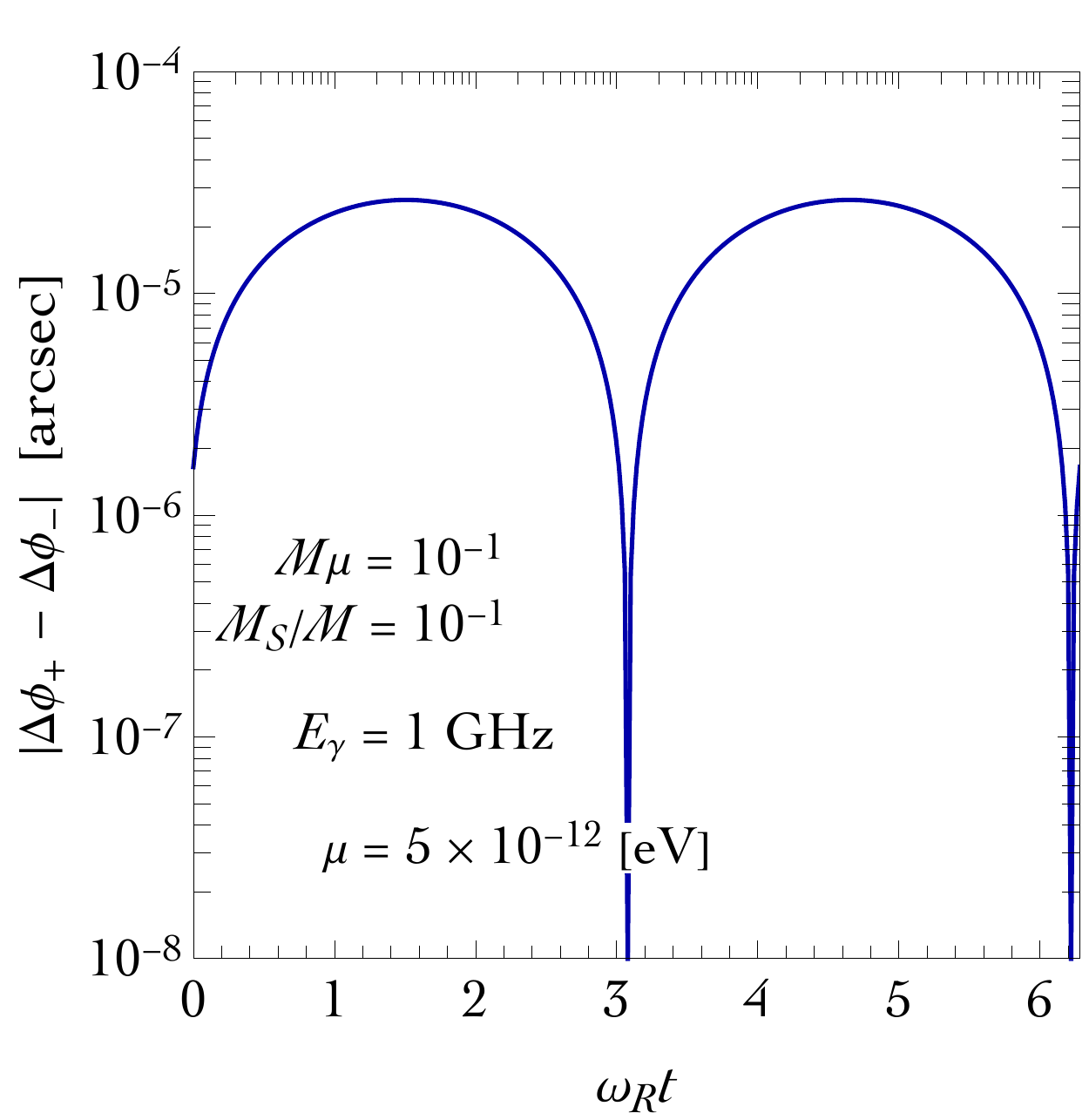}
\endminipage 
\vspace{-.2 cm}
\caption{\label{fig:QCDBending}\em 
Left panel. Contours of constant angular splitting $|\Delta\phi_+ - \Delta\phi_-|$ (for fixed time $t$) as a function 
of the radio wave energy $E_{\gamma}$ and the QCD axion mass $\mu$. 
We fix $M\mu = M_S/M = 10^{-1}$, and we explore different possibilities for the 
 electromagnetic-to-color anomaly ratio $E/N$ in eq.~(\ref{eq:AxionDecayConstant}). Right panel.
  Time-dependence of the angular splitting $|\Delta\phi_+ - \Delta\phi_-|$ for fixed QCD axion mass and radio wave energy. The period of the signal is set by the inverse of $\omega_{R}\approx \mu$, and 
  we have $1/\mu \approx 0.66\times 10^{-3}\,(10^{-12}\,{\rm eV}/\mu)$ s.
}
\end{figure}
We show our result in fig.~\ref{fig:QCDBending}. 
We imagine a ray of light with energy $E_{\gamma}$ traveling through the axion cloud, and 
in the left panel
we plot (at fixed $t$) the angular splitting $|\Delta\phi_+ - \Delta\phi_-|$ as a function of $E_{\gamma}$ 
and the axion mass $\mu$. We fix $M\mu = M_S/M = 10^{-1}$, and we consider different values 
for the parameter $E/N$ in eq.~(\ref{eq:AxionDecayConstant}). 
Since $M\mu$ is fixed, at each value of $\mu$ corresponds a black hole mass $M$ (respectively, left and right y axis). As expected, the QCD axion is relevant 
 in connection with stellar-mass black holes.
For typical values $0 < E/N < 8/3$, we obtain an angular splitting between left and right polarization of the order $10^{-7} < |\Delta\phi_+ - \Delta\phi_-| [{\rm arcsec}] < 10^{-9}$. As we shall discuss in section~\ref{sec:Discussion}, these values are probably too small for a detection since, even taking an optimistic view, it is not possible at present to reach angular resolutions
 below $\delta\theta \approx 10^{-6}$ arcsec.
  For the QCD axion $|\Delta\phi_+ - \Delta\phi_-| \simeq 10^{-6}$ arcsec can be obtained in tha analyzed parameter space
for $E/N = 170/3$ (dot-dashed black lines in fig.~\ref{fig:QCDBending}).

In the right panel of fig.~\ref{fig:QCDBending} we show the time-dependence of 
$|\Delta\phi_+ - \Delta\phi_-|$ due to the rotation of the cloud. 
 We choose $\mu = 5\times 10^{-12}$ eV and fixed energy $E_{\gamma} = 1$ GHz. 
  The signal displays the expected periodicity set by 
  $T = 2\pi/\omega_{\rm R}\simeq 2\pi/\mu$. 
  
As far as the QCD axion is concerned, 
the relevance of the polarization-dependent bending  seems to be quite modest.
The reason is that eq.~(\ref{eq:AxionDecayConstant}) and eq.~(\ref{eq:QCD}) imply a very strong relation 
 between the mass of the QCD axion and its coupling to photons, and the range 
 explored in eq.~(\ref{eq:GoldenRange}) corresponds to a coupling $g_{a\gamma\gamma}$ that is too weak.
  However, this is not a lapidary conclusion. 
  The way-out is that 
 the  relation between the axion mass and the axion-photon coupling 
  can not be considered  a solid prediction of QCD, in clear contrast with the relation between axion mass and axion decay constant.
 The latter is dictated by the minimization of the effective potential generated by 
  the explicit breaking of the continuous global shift symmetry of the axion due to QCD instanton effects, and thus tightly linked to the solution of the strong CP problem. 
 The former has a degree of model-dependence -- a fact already clear from the discussion about the possible values of $E/N$ below eq.~(\ref{eq:AxionDecayConstant}) -- that can be exploited.
It is possible, therefore, to construct simple models in which 
the axion-photon coupling can be arbitrarily large without altering eq.~(\ref{eq:QCD}). 
In the next section, we shall illustrate one explicit realization of this idea.

\subsection{The photo-philic QCD axion}\label{sec:PhotoPhilic}

The photo-philic ($\gamma_\heart$ hereafter) QCD axion~\cite{Farina:2016tgd} 
 is a specific realization of the clockwork mechanism proposed in~\cite{Choi:2015fiu,Kaplan:2015fuy}.
 In its original incarnation, the clockwork is a renormalizable theory that consists in a chain of $\mathcal{N}+1$ complex scalar fields with a $U(1)^{\mathcal{N}+1}$ global symmetry 
 spontaneously broken at the scale $f$. The $U(1)^{\mathcal{N}+1}$ global symmetry is also 
  explicitly broken 
in such a way to preserve a single $U(1)$ symmetry whose 
 Nambu--Goldstone boson -- eventually identified with the QCD axion in~\cite{Farina:2016tgd} -- lives in a compact field space with a dimension that is set by the effective decay constant 
  $f_{a} = 3^\mathcal{N} f \gg f$.
  The key idea of~\cite{Farina:2016tgd} is the following.
 New vector-like fermions which are responsible 
for the generation of the color anomaly are coupled to the last site $\mathcal{N}$ of the scalar chain. 
This guarantees the usual solution of the strong CP problem with the important difference that 
 the scale $f_a = 3^\mathcal{N} f$ entering in eq.~(\ref{eq:QCD}) can be parametrically much larger 
 than the fundamental symmetry breaking scale $f$. This feature has very important phenomenological consequences because the model 
 predicts the presence of additional pseudo-scalar particles which can be light and accessible at the LHC 
  while keeping $f_a$ above the astrophysical bounds (roughly $f_a \gtrsim 10^9$ GeV). 
  In the usual realization of the QCD axion presented in section~\ref{sec:QCDaxion}, 
   the same vector-like fermions mediating the QCD anomaly 
    also contribute to the axion-photon coupling. 
    In the $\gamma_\heart$ QCD axion, on the contrary, there are additional electromagnetically charged 
    vector-like fermions coupled to the site $\mathcal{M}<\mathcal{N}$ of the scalar chain. These fermions are responsible for 
    the axion-photon coupling that is, by all accounts, 
    disentangled from the solution of the strong CP problem. In the simplest  
     realization proposed in~\cite{Farina:2016tgd}, 
      the $\gamma_\heart$ QCD model requires the existence of a single pair of vector-like colored 
       fermions in the fundamental representation of $SU(3)_C$ and 
 a single pair of color neutral vector-like fermions with unit hypercharge  and singlet under $SU(2)_L$.
  Under these conditions the axion-photon coupling turns out to be~\cite{Farina:2016tgd}
 \begin{equation}\label{eq:PappaMain}
  g_{a\gamma\gamma} = \left(\frac{2}{3^{\mathcal{M}-\mathcal{N}}}\right)\frac{\alpha_{\rm em}}{2\pi f_a}~,
  \end{equation} 
  and the free parameter $\mathcal{N}$, that is a fundamental parameter of the model, can be changed 
  to make $g_{a\gamma\gamma}$, as promised, arbitrarily large.  
\begin{figure}[!htb!]
  \begin{center}
\fbox{$\gamma_\heart$ QCD axion (right) and axion-like particles (left)}\vspace{-0.1cm}
\end{center}
\minipage{0.5\textwidth}
\centering
  \includegraphics[width=.9\linewidth]{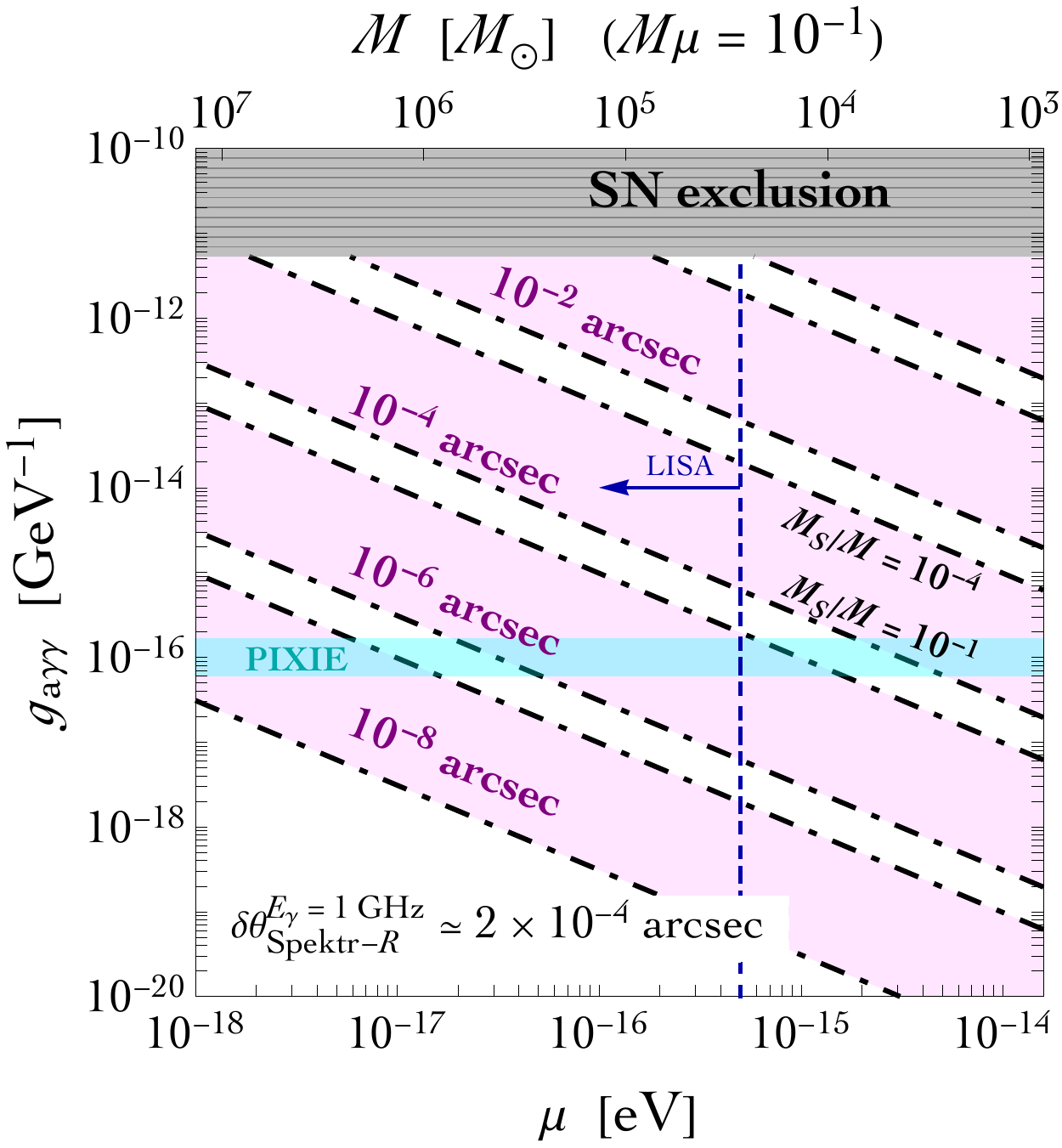}
\endminipage 
\minipage{0.5\textwidth}
  \includegraphics[width=.9\linewidth]{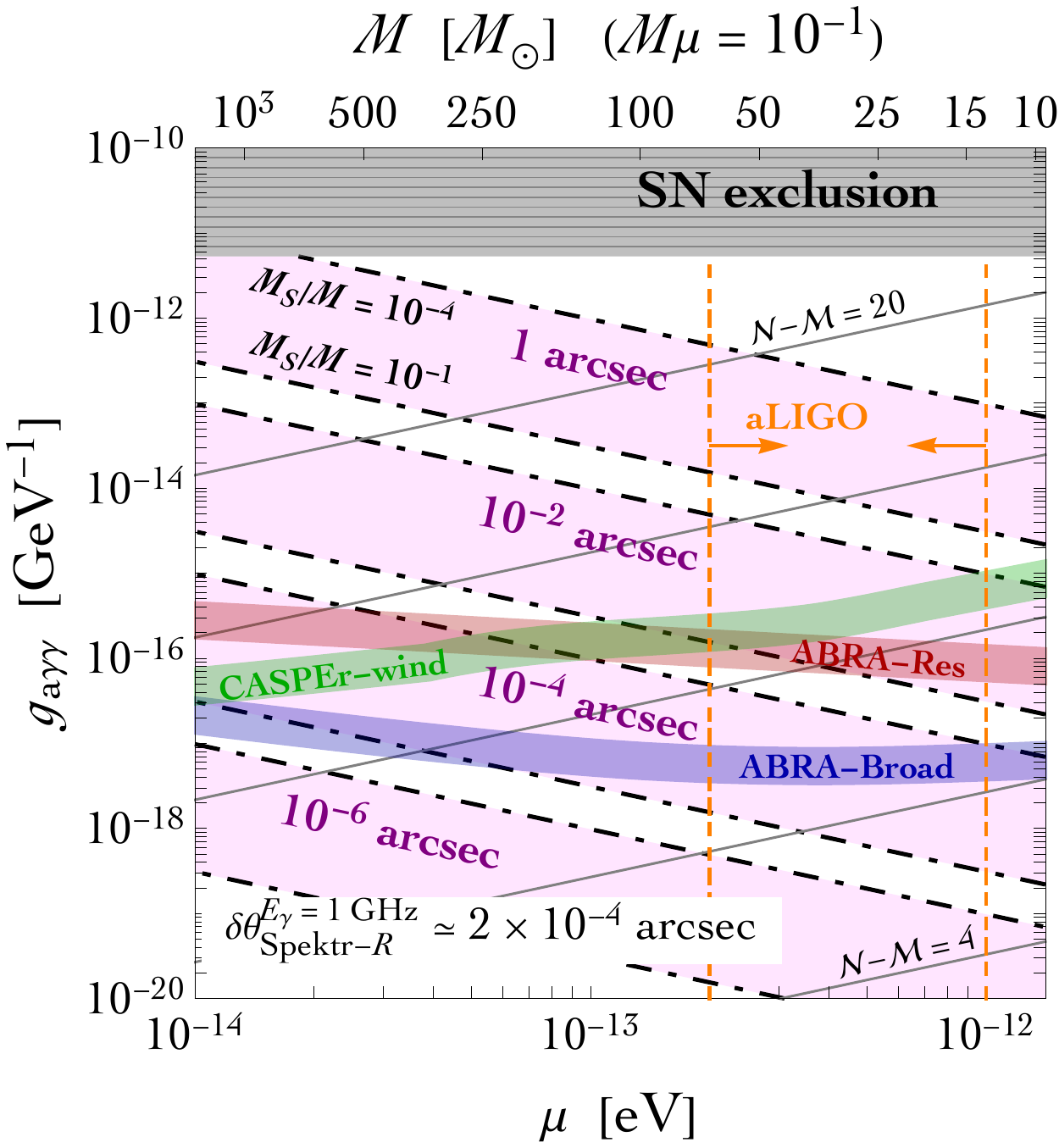}
\endminipage 
\vspace{-.2 cm}
\caption{\label{fig:ALPsParameterSpace}\em 
Right (left) panel. Contours of constant angular splitting for the $\gamma_\heart$ QCD axion (a generic ALP) as a function of the axion mass $\mu$ and the axion-photon coupling $g_{a\gamma\gamma}$. 
In the case of the $\gamma_\heart$ QCD axion we show the projected  sensitivities of ABRACADABRA~\cite{Kahn:2016aff} 
and CASPEr-wind~\cite{Graham:2013gfa} together with the mass range that will be explored by the Advanced LIGO gravitational wave interferometer~\cite{Brito:2017wnc,Brito:2017zvb}. 
For a generic ALP, we show a projected limit for the PIXIE/PRISM experiment~\cite{Tashiro:2013yea} (see text for details). 
}
\end{figure}
   
In the right panel of fig.~\ref{fig:ALPsParameterSpace} we show the result of our analysis 
for the $\gamma_\heart$ QCD axion. 
We explore the parameter space $(\mu,\,g_{a\gamma\gamma})$, and we fix $M\mu = 10^{-1}$.
 We enlarge the axion mass range to the interval $10^{-14} \leqslant \mu~[{\rm eV}] \leqslant 10^{-12}$, and we bracket between two vertical dot-dashed orange lines the mass range covered by Advanced LIGO in eq.~(\ref{eq:GoldenRange}). The above mass range corresponds to the axion decay constant 
 $5.7 \times 10^{18} \lesssim f_a~[{\rm GeV}]\lesssim 5.7\times 10^{20}$, and in order to reproduce the observed value of the dark matter relic abundance we need to tune the initial misalignment angle to the values 
 $0.12 \lesssim \theta_{\rm in}\times 10^5~[{\rm rad}] \lesssim 3.98$. 
 We consider the axion-photon coupling in the range $10^{-20}\leqslant g_{a\gamma\gamma}~[{\rm GeV}^{-1}] \leqslant 10^{-10}$, and the thin diagonal solid gray lines indicate -- in steps of $4$, from $\mathcal{N}-\mathcal{M} = 4$ to $\mathcal{N}-\mathcal{M} = 20$ -- the values of $g_{a\gamma\gamma}$ as a function of the axion mass for different choices of $\mathcal{N}-\mathcal{M}$ in eq.~(\ref{eq:PappaMain}).
 Contours of constant angle  $|\Delta\phi_+ - \Delta\phi_-|$ are shown with dot-dashed diagonal black lines, and the shaded area in magenta corresponds to $10^{-4} \leqslant M_S/M \leqslant 10^{-1}$. 
 We fix $E_{\gamma} = 1$ GHz, and -- to give an idea about the relevance of the effect -- we quote the angular resolution of the Spektr-R radio telescope~\cite{Kardashev:2013cla,RadioAstronWeb,RadioAstron}, 
 $\delta\theta_{\rm Spektr-R}^{E_{\gamma} = 1\,{\rm GHz}} \simeq 2\times 10^{-4}$ arcsec. 
 We postpone to section~\ref{sec:Discussion} a more detailed discussion about experimental prospects.
The gray area is excluded by SN1987A gamma-ray limit on ultralight axion-like particles, and we use the 
results of the updated analysis presented in~\cite{Payez:2014xsa}. 
 The plot shows that $|\Delta\phi_+ - \Delta\phi_-| > \delta\theta_{\rm Spektr-R}^{E_{\gamma} = 1\,{\rm GHz}}$
in a wide range of the explored parameter space. 
We argue that the polarization-dependent lensing computed  in 
 section~\ref{sec:Bending} can be relevant for the $\gamma_\heart$ QCD axion.
It is also important to keep in mind that the same region of parameter space is well within the sensitivity range  of well-motivated proposals for future experiments.
In the right panel of fig.~\ref{fig:ALPsParameterSpace} we show the 
  projected  sensitivities of ABRACADABRA~\cite{Kahn:2016aff} (considering both the resonant and broadband approach)
and CASPEr-wind~\cite{Graham:2013gfa}.
ABRACADABRA exploits the fact that when axion dark matter encounters a static magnetic field, it sources an effective electric current that follows the magnetic field lines and oscillates 
at the axion Compton frequency. CASPEr-wind considers couplings of 
the background classical axion field which give rise to observable effects like
 nuclear electric dipole moment, and axial nucleon and electron moments.

\subsection{Axion-like particles}

We now turn to discuss the more general case of ALPs.
 The crucial difference
   is that there is no {\it a priori} relationship between 
   the ALP mass $\mu$ and the 
  coupling $g_{a\gamma\gamma}$
 while in the QCD axion case they are linearly related, and we can therefore treat them as
  independent parameters. As a result, ultra-light values of $\mu$ below those
  explored in section~\ref{sec:QCDaxion} and~\ref{sec:PhotoPhilic} are possible.
We show our result in the left panel of fig.~\ref{fig:ALPsParameterSpace}.
In order to provide complementary information with respect to the case  of the 
$\gamma_\heart$ QCD axion,   we consider the mass range $10^{-18} \leqslant \mu~[{\rm eV}] \leqslant 10^{-14}$. Since $M\mu =10^{-1}$, this range covers from intermediate-mass to supermassive black holes.
As far as the computation of $|\Delta\phi_+ - \Delta\phi_-|$ is concerned, 
the color code follows what already discussed in section~\ref{sec:PhotoPhilic}. 
We delimit with a vertical dot-dashed blue line the mass range that will be explored by LISA according to the analysis proposed in~\cite{Brito:2017wnc,Brito:2017zvb}.
We find that $|\Delta\phi_+ - \Delta\phi_-| > \delta\theta_{\rm Spektr-R}^{E_{\gamma} = 1\,{\rm GHz}}$
in a wide range of the explored parameter space, and we argue that the polarization-dependent 
effect computed in section~\ref{sec:Bending} can be relevant also for a generic ALP.
We also show a possible complementarity with future CMB tests of dark matter.
The idea is that 
resonant conversions between CMB photons and light ALPs could result in observable CMB distortions.
These resonant conversions depend on the strength of primordial magnetic fields $B$,
 and it was shown in~\cite{Tashiro:2013yea} that
 the PIXIE/PRISM experiment~\cite{Andre:2013afa}, according to the expected sensitivity, has the capabilities to set the limit  
 $g_{a\gamma\gamma}B \lesssim 10^{-16}$ GeV$^{-1}$ nG for axion mass $\mu \lesssim 10^{-14}$ eV (see also~\cite{Schlederer:2015jwa} for a recent analysis using galaxy clusters).  
Assuming a strength of primordial magnetic fields close to the current upper limit $B \sim \mathcal{O}(1)$ nG~\cite{Ade:2015cva}, we show in cyan the expected 
limit on $g_{a\gamma\gamma}$ in  fig.~\ref{fig:ALPsParameterSpace}.

\section{Discussion and outlook}\label{sec:Discussion}

\begin{figure}[!htb!]
\centering
  \includegraphics[width=.95\linewidth]{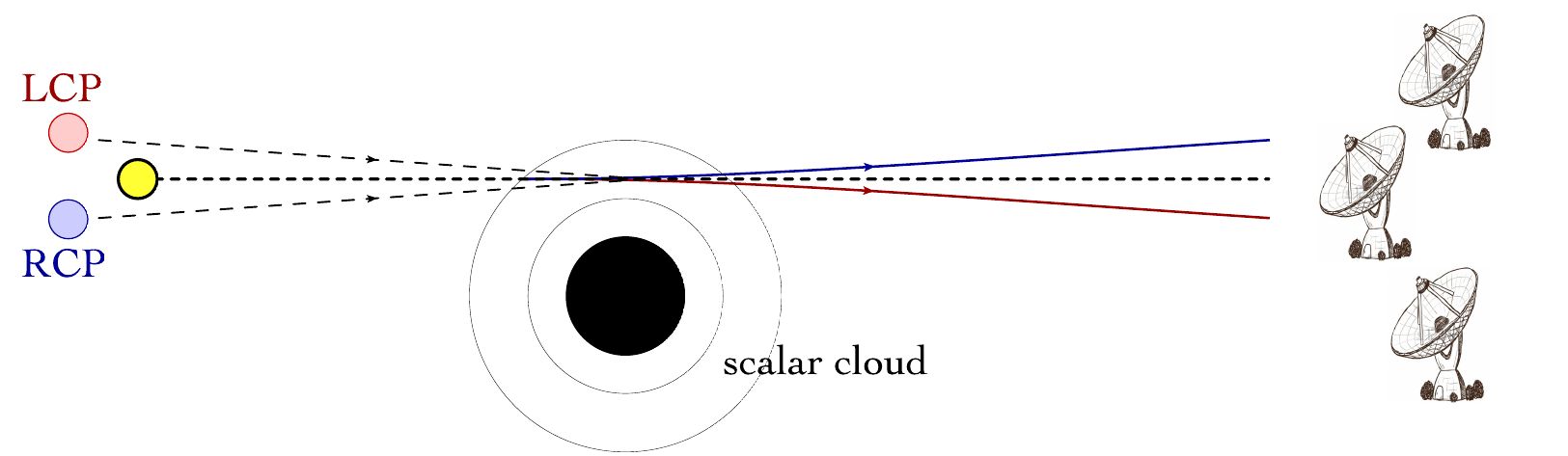}
\vspace{-.2 cm}
\caption{\label{fig:SetUp}\em 
Sketch of the typical configuration needed to detect the polarization-dependent bending discussed in 
section~\ref{sec:Bending}. An astrophysical radio source like an AGN emits linearly polarized light.
 Traveling through the axion scalar cloud surrounding  a Kerr black hole, 
 the left and right circular components (LCP and RCP) experience different deflection angles
 thus creating a polarization-dependent lensing that could be observed 
 by  an array of radio telescopes using the VLBI technique.
}
\end{figure}

The setup we have in mind is sketched  in fig.~\ref{fig:SetUp}. 
We envisage the presence of a black hole surrounded by a scalar cloud in between 
an astrophysical source emitting linearly polarized light and a ground- or space-based radio telescope.
A linearly polarized ray of light is a superposition of right- and left-handed circularly polarized waves (RCP and LCP in fig.~\ref{fig:SetUp}).
By traveling trough the scalar cloud, the two components experience a polarization-dependent bending as discussed in the previous section and appendix~\ref{app:C}. 
In that event, a polarization-dependent lensing effect would appear in the image captured by the radio telescope.

Is this situation ever possible? In this section, we shall explore in more detail 
some of the necessary conditions needed to realize
 this idea.

\subsection{General remarks: dual-polarization receiver and VLBI}

Consider an electromagnetic wave traveling in the $\hat{z}$ direction.
 In general, light is elliptically polarized and can be described by means of the electric field
\begin{equation}
\vec{E}_{\rm EP} =
E_x^{(0)}\cos(kz -\omega t)\hat{x} + E_y^{(0)}\cos(kz -\omega t + \delta)\hat{y}
\equiv E_x \hat{x} + E_y \hat{y}~.
\end{equation}
The case $\delta = 0$ corresponds to linear polarization 
whereas
 the conditions $\delta = \pm \pi/2$, $E_x^{(0)} = E_y^{(0)}$
describe, respectively, a right and left circularly polarized wave.
 What is relevant for astrophysical observations is light intensity 
 rather than field amplitude. 
 For this reason it is useful to introduce the four Stokes parameters~\cite{RadioBook}
 \begin{equation}\label{eq:Stokes}
 \textgoth{I} = \langle E_x^2\rangle + \langle E_y^2\rangle~,~~~\textgoth{Q} = \langle E_x^2\rangle - \langle E_y^2\rangle~,~~~\textgoth{U}=2\langle E_x E_y\cos\delta\rangle~,~~~
 \textgoth{V}=2\langle E_x E_y\sin\delta\rangle~,
 \end{equation}
 where $\langle \cdots \rangle$ denotes a time average over times much larger than $2\pi/\omega$.
  The parameter $\textgoth{I}$ measures the intensity of the wave, $\textgoth{Q}$
  and $\textgoth{U}$ fully describe linear polarization, and 
     $\textgoth{V}$ corresponds to circularly polarized intensity. 
     In particular, a net right (left) polarization has a positive (negative) $\textgoth{V}$.
 
The radio emission from most bright radio sources arises from synchrotron radiation, and it is linearly polarized.
Qualitatively speaking, the reason is the following. 
The radiation from a single relativistic electron gyrating around a magnetic field is 
elliptically polarized. For an ensemble of electrons with a smooth distribution of pitch 
angles the opposite senses of elliptical polarization will cancel out resulting in 
linearly polarized radiation. 
This is in particular true in the case of synchrotron emission from Active Galactic Nuclei (AGN) observed at radio frequencies. This is, therefore, the class of astrophysical sources that might be well-suited for our purposes.

Next, we need a radio telescope able to distinguish between
 left and right polarizations with sufficiently high angular resolution. 
 Polarization-dependent measurements are possible if the instrument is a dual-polarization receiver. 
 In a nutshell, such telescope can be thought as a cross of two dipoles aligned along orthogonal directions.
 Each of the two dipoles measures the corresponding polarization component and converts it into an electric signal. The signals are auto-correlated and cross-correlated, thus allowing for a reconstruction of the Stokes parameters. 
What is important to stress is that all four Stokes parameters are actual intensities.
 This means that 
 they can be  
 used at the level of image analysis in order to reconstruct and visualize  the polarization 
 of the observed source. This makes the detection of our effect, at least in principle, possible.
  Furthermore, we remind that the time average implied in the measurement of the Stokes parameters 
  refers to a time interval $\Delta t$ much larger than the typical wavelength $\lambda$ of the observed light.
  If the condition $\lambda\ll \Delta t \ll \lambda_{\rm Compton}$ is verified, 
   it could even be possible to detect the time variation of the signal.

Let us now comment about the angular resolution.
The angular resolution $\delta\theta$ of a telescope can be calculated from the wavelength of observed 
 radio waves $\lambda$ and the diameter $D$ of the telescope
 \begin{equation}
 \delta\theta \approx 2.5\times 10^{5}~\frac{\lambda}{D}~{\rm arcsec}~.
 \end{equation}
 To fix ideas, a radio telescope with $D = 65$ m 
 observing radio wavelengths at $E_{\gamma} = 1$ GHz ($\lambda \approx 0.3$ m)
  has an angular resolution $\delta\theta \approx 10^3$ arcsec.
  The angular resolution of 
  a typical radio telescope is, therefore, by far too low to detect the effect computed in section~\ref{sec:Bending}.
  However, it is possible to use multiple radio telescopes at the same time, a technique that is called interferometry. The angular resolution in greatly improved 
  because -- by synchronizing and combining observations from all the telescopes of the array, each one equipped by its own atomic clock -- 
   one effectively creates a single telescope as large as the distance between the two farthest telescopes.
   This simple principle lies at the heart of the 
   very-long-baseline interferometry (VLBI) technique, in which a signal from an astronomical 
   radio source is collected from multiple radio telescopes on Earth.
VLBI gives angular resolutions of the order of 
$\delta\theta \approx 10^{-3}$ arcsec or better 
thus making our speculations about a possible detection more realistic.
 
 A further improvement can be obtained by combining a VLBI array with an additional 
 antenna placed  on board of a satellite orbiting the Earth.
As a benchmark reference, let us consider the case of the Russian project
 Spektr-R~\cite{Kardashev:2013cla,RadioAstronWeb,RadioAstron}.
Spektr-R (formerly RadioAstron) is a dual-polarization receiver space-based 10 meter radio telescope 
in a highly apogee orbit around the Earth, launched on July 2011.
Spektr-R works  in conjunction with some of the largest ground-based radio telescopes, and 
the system forms an interferometric baseline extending up to $3\times 10^5$ 
km~\cite{Kardashev:2013cla,RadioAstronWeb,RadioAstron}. 
 This configuration is able to reach an astonishing 
 angular resolution up to a few millionths of an arcsecond.
As a reference, in fig.~\ref{fig:ALPsParameterSpace} we quote the typical angular resolution 
of Spektr-R at $E_{\gamma} = 1$ GHz, that is about $\delta\theta \approx 2\times 10^{-4}$ arcsec.

In conclusion, we argue that radio astronomy techniques 
 have the capabilities to detect the polarization-dependent bending discussed in section~\ref{sec:Bending}, if realized in Nature.
  Of course,
  for the aim of the present work our discussion is purely qualitative, 
  and our intent is that of stimulating the interplay with 
 the radio astronomy community to fully understand the validity of 
  our conclusions.

\subsection{Comparison with ``background'' effects}

Scintillation is an optical effect arising when light rays emitted by a compact source pass through a turbulent ionized 
medium. 
As far as radio frequencies are concerned, 
scintillation theory can be applied to the turbulent interstellar medium (ISM) of the Galaxy 
through angular and pulse broadening of pulsars~\cite{Taylor:1993my,Armstrong:1995zc,Cordes:2002wz}, 
and
to the turbulent intergalactic medium (IGM) through quasar observations~\cite{Sciamano,Ferrara:2001ib,Pallottini:2013rja}.

Interstellar scattering of an extragalactic source of radio waves results in angular broadening.
It is, therefore, important to keep in mind the typical size of this effect since it acts  
 as a sort of ``background'' for the polarization-dependent effect discussed in section~\ref{sec:Bending}. 
  If the angular broadening proves to be much larger than the angular splitting 
   $|\Delta\phi_+ - \Delta\phi_-|$, we expect the latter to be clouded by the former.

The size of the broadening of an extragalactic source at redshift $z_S$ due to the IGM  -- modeled as a thin-screen at redshift $z_L$ with homogeneous 
Kolmogorov turbulence -- is~\cite{Macquart:2013nba}
\begin{equation}\label{eq:ThetaScatt}
\theta_{\rm scat} \sim 19.75\,\,{\rm SM}^{3/5}\,\left(
\frac{D_{\rm LS}}{D_{\rm S}}
\right)\left(
\frac{E_{\gamma}}{1\,{\rm GHz}}
\right)^{-2.2}(1+z_L)^{-1.2}\,10^{-3}\,{\rm arcsec}~,
\end{equation}
where $D_{\rm LS}$ ($D_{\rm S}$) is the angular diameter distance between the scattering region and the source (between the observer and the source).
The angular diameter distance at redshift $z$ is given by the integral
\begin{equation}
D(z) = cH_0^{-1}(1+z)^{-1}\int_0^z\left[
\Omega_{\Lambda} + (1-\Omega)(1+z^{\prime})^2 + \Omega_m(1+z^{\prime})^3 + \Omega_r(1+z^{\prime})^4
\right]^{-1/2}dz^{\prime}~,
\end{equation}
where $H_0$ is the Hubble constant, $\Omega = \Omega_{\Lambda} + \Omega_m +  \Omega_r$, and 
$\Omega_{\Lambda}$,  $\Omega_m$,  $\Omega_r$ are, respectively, the ratios of the dark energy density, matter density and radiation density to the critical density of the Universe.
We assume Standard Cosmology, with $\Omega = 1$, $\Omega_{\Lambda} = 0.7$, and $\Omega_r = 0$. In eq.~(\ref{eq:ThetaScatt}) we introduced
 the short-hand notation $D(z_i) \equiv D_i$.
We use $H_0 = 67.8\pm 0.9$ (km/s)/Mpc~\cite{Ade:2015xua}. 
\begin{figure}[!htb!]
\begin{center}
  \includegraphics[width=.5\linewidth]{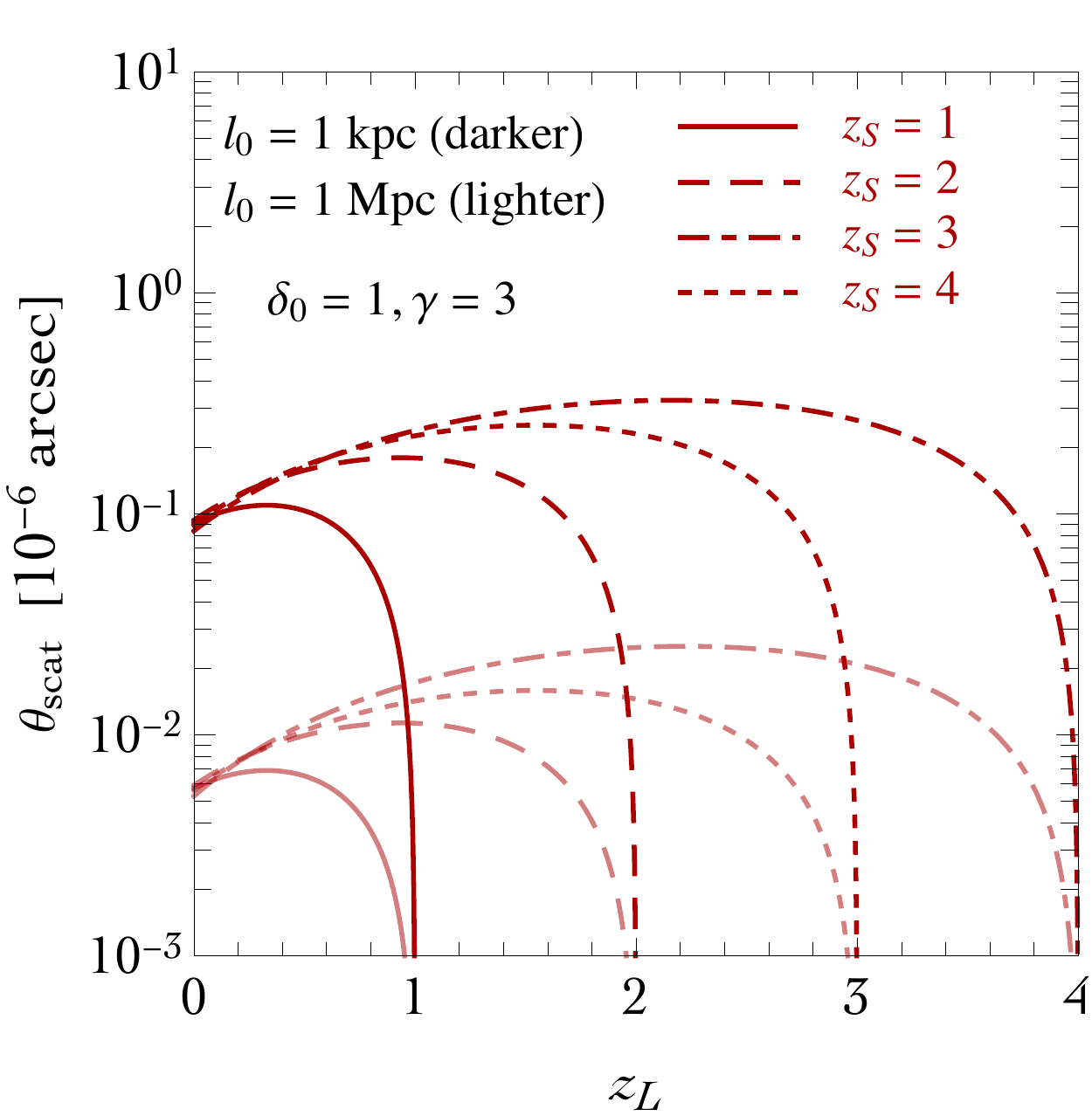}
\end{center}
\vspace{-.25 cm}
\caption{\label{fig:IGM}\em 
Angular broadening in the IGM at $E_{\gamma} = 1$ GHz for a screen at redshift $z_L$.
}
\end{figure}
Notice that -- defining the angular diameter distance between the observer and the scattering region as $D_{\rm L}$ --  we have 
in general $D_{\rm LS} \neq D_{\rm S} - D_{\rm L}$. 
In eq.~(\ref{eq:ThetaScatt}), the scattering measure ${\rm SM}$ encodes the level of turbulence of the IGM, and can be defined as 
the line-of-sight integral of the spectral coefficient characterizing the power spectrum of electron density fluctuations.
Following~\cite{Macquart:2013nba}, we have 
${\rm SM} = C_{\rm SM}\,F\,n_e^2(z)\,D_S$. The constant $C_{\rm SM}$ takes the value $C_{\rm SM} = 1.8\,{\rm m}^{-20/3}\,{\rm cm}^6$, $n_e(z)$ is the electron density at redshift $z$, and the fluctuation 
parameter is $F = (\zeta \epsilon^2/\eta)(l_0/1\,{\rm pc})^{-2/3}$~\cite{Taylor:1993my} where 
$l_0$ is the outer scale of the turbulence, $\eta$ is the filling factor of the turbulent medium, 
$\epsilon$ is the variance of the electron density fluctuations within a single cloud, 
and $\zeta$ is a measure of fluctuations in the mean density between clouds.
We assume in our estimate $\epsilon \sim \zeta \sim \eta \sim 1$ for all redshifts. This choice implies that the turbulence is fully developed at all redshifts of interest.
The outer scale length of turbulence $l_0$ defines an upper cut off in the size of turbulent structures, and we consider the two benchmark values $l_0 = 1$ kpc,  $l_0 = 1$ Mpc.  
The mean free electron density as a function of the redshift is given by 
$n_e(z) = \delta_0\,x_e(z)\,n_e(0)\,(1+z)^{\gamma}$, where $x_e(z)$ is the ionization fraction,  and $n_e(0) = 2.1\times 10^{-7}$ cm$^{-3}$ 
is the mean free electron density at $z = 0$. We assume a significant ionized fraction,  $x_e(z) \sim 1$, for all redshifts of interest. 
The parameter $\delta_0$ controls possible electron overdensity while $\gamma \sim 3$ for IGM components with constant comoving densities.
For simplicity, we take $\delta_0 = 1$. The presence of possible electron overdensity results in a rescaling of eq.~(\ref{eq:ThetaScatt}) according to the factor $\delta_0^{6/5}$.
In fig.~\ref{fig:IGM} we show the angular broadening predicted by eq.~(\ref{eq:ThetaScatt}) 
at $E_{\gamma} = 1$ GHz for a screen of ionized medium at redshift $z_L$. We consider four different source locations, at $z_S = 1,2,3,4$, and two possible choices for the 
outer scale of the turbulence $l_0$ (see caption for details).
The scattering angle ranges between $10^{-9} \lesssim \theta_{\rm scat}  \lesssim10^{-7}$ arcsec for $1\,{\rm kpc} \lesssim l_0 \lesssim 1\,{\rm Mpc}$.
We notice that the scattering broadening in the medium hosted by the background source (i.e. considering scattering screens located at 
$z_L\simeq z_S$) drops to negligible values. Finally, changing the spectral index $\gamma$ results in a different $z_L$ dependence of 
the scattering angle, but it does not alter the order of magnitude estimate of the broadening effect.

Given the model-dependence and the astrophysical uncertainties entering in the computation of the angular broadening, no firm conclusion can be established.
  Nevertheless, the order-of-magnitude estimate proposed in this section  
  keeps alive the hope of detecting the polarization-dependent bending due to a superradiant axion cloud.

\subsection{Faraday rotation}

Finally, let us close this section with a short discussion about another important effect that is usually  relevant in the presence of an optically active medium: Faraday rotation.

Consider a beam of light linearly polarized along the $\hat{x}$ axes
\begin{equation}
\vec{E}_{\rm LP} = E_0\cos(kz-\omega t)\hat{x}~,~~~~~~~~~{\rm with}~~~k=2\pi/\lambda~,~\omega = 2\pi\nu~.
\end{equation}
A linearly-polarized wave can be decomposed into a sum of left- and right-circularly polarized waves at the same frequency
\begin{equation}
\vec{E}_{\rm LP} = \frac{\vec{E}_{\rm RCP} + \vec{E}_{\rm LCP}}{2}~,~~~~{\rm with}~~~
\vec{E}_{\rm RCP,LCP} = E_0\left[ 
\cos(kz-\omega t)\hat{x} \pm \sin(kz-\omega t)\hat{y}
\right]~.
\end{equation}
Imagine this beam enters a region characterized by the presence of a medium which has slightly different propagation velocities 
for light with opposite circular polarizations.
Upon exiting this region, the left- and right-circular polarization modes
 have picked up a net phase difference
 \begin{equation}
 \vec{E}_{\rm RCP,LCP} = E_0\left[ 
\cos(kz-\omega t + \delta_{\rm R,L})\hat{x} \pm \sin(kz-\omega t + \delta_{\rm R,L})\hat{y}
\right]
 \end{equation}
  which causes their sum to still be linearly-polarized, but along a different axis.
   Indeed the sum $\vec{E}_{\rm LP} = (\vec{E}_{\rm RCP} + \vec{E}_{\rm LCP})/2$
\begin{equation}
\vec{E}_{\rm LP} = E_0\left[
\cos\left(
\frac{\delta_{\rm R} - \delta_{\rm L}}{2}
\right)\hat{x} + 
\sin\left(
\frac{\delta_{\rm R} - \delta_{\rm L}}{2}
\right)\hat{y}
\right]\cos\left(
kz -\omega t + \frac{\delta_{\rm R} + \delta_{\rm L}}{2}
\right)~,
\end{equation}
 describes a plane polarized wave with the polarization direction twisted by an angle 
 $\Delta \equiv (\delta_{\rm R} - \delta_{\rm L})/2$ from the $x$-axis towards the $y$-axis.
  This is the Faraday rotation.
  
  The parity violating interaction in eq.~(\ref{eq:AxionPhotonL}) may induce Faraday rotation for a beam of light traveling through the axion cloud. 
We can estimate the size of such effect by considering 
a wave traveling a distance $L\sim r_{\rm cloud}$ in the equatorial plane
at radial distance $r \sim r_{\rm max}$.  
The change in phase of a circularly polarized mode traveling a 
distance $L$ is $\delta = L|\vec{k}|$. From eq.~(\ref{eq:DispApp}), and considering the approximation 
discussed 
in eq.~(\ref{eq:approximation}), at the linear order in $g_{a\gamma\gamma}$ we have
 $|\vec{k}| \approx E_{\gamma} \mp (g_{a\gamma\gamma}/2)\partial\Phi/\partial t$.
  We therefore find the estimate $\Delta = L(g_{a\gamma\gamma}/2)\left.\partial\Phi/\partial t\right|_{r = r_{\rm max}}$ where 
  for simplicity we 
  assumed a constant cloud (with value fixed at $r = r_{\rm max}$) along the distance $L$.
   We also neglected the trigonometric factor that is responsible for the rotation of the cloud.
   This estimate should be therefore considered as an order-of magnitude upper limit for the effect.
For the QCD axion and for a generic ALP we find
\begin{eqnarray}
\Delta_{\rm QCD} &=& 2\times 10^{-5}\left(\frac{E}{N} - 1.92\right)
\left(
\frac{\mu}{10^{-12}\,{\rm eV}}
\right)\left(
\frac{M_S/M}{0.1}
\right)^{1/2}\left(
\frac{M\mu}{0.1}
\right)~{\rm rad}~,\label{eq:Faraday1}\\
\Delta_{\rm ALP} &=& 10\left(
\frac{g_{a\gamma\gamma}}{10^{-16}\,{\rm GeV}^{-1}}
\right)
\left(
\frac{M_S/M}{0.1}
\right)^{1/2}\left(
\frac{M\mu}{0.1}
\right)~{\rm rad}~.\label{eq:Faraday2}
\end{eqnarray}
Our Galaxy is full of ionized hot gas, and is
 simultaneously permeated by a large-scale magnetic field. 
 The Faraday effect due to this plasma is observed in the polarized signal from radio pulsars within our Galaxy, and on all extragalactic radio sources. The subtlety is that 
 we do not know the original plane of polarization. As a consequence, 
 the effect is almost always studied as a function of frequency. 
 In this case the Faraday rotation has the simple form $\Delta = {\rm RM}\,\lambda^2$,  
 where $\lambda$ is the wavelength of the observed light and RM is the  rotation measure which 
 in general depends on the interstellar magnetic field and the number density of electrons along the propagation path. 
  In the idealized case, one
can determine the RM by measuring $\Delta$ at different wavelengths, and then performing a linear fit. 
From the value of RM, one can in turn try to decrypt  the physical conditions along the lines of sight.

 The effect proposed in eqs.~(\ref{eq:Faraday1},\,\ref{eq:Faraday2}) does not feature any energy
  dependence. 
  Without knowing the original direction of polarization, therefore, 
  a possible detection of this effect seems hopeless.
   One possibility is to exploit the time-dependence of the signal, similar to the one discussed in the right panel of fig.~\ref{fig:QCDBending}, that should give rise to a time-dependent oscillating effect with period set by $1/\mu$.
   
Another interesting aspect is to consider as a source of light the accretion disk surrounding 
the black hole (instead of a distant source as done in section~\ref{sec:Discussion}).
 Gravitational and frictional forces compress and raise the temperature of the material in the disk, 
 thus causing the emission of electromagnetic radiation that should travel through the axion cloud before escaping. 

We do not explore further such possibilities, 
and we postpone a more detailed investigation to future work.

\section{Conclusions}\label{sec:Conclusions}

Black holes were long considered a mathematical curiosity rather than a true prediction of General Relativity realized in Nature.
After the 
 first direct detection of gravitational waves and the first observation of a binary black hole 
 merger~\cite{Abbott:2016blz},
the possibility to turn black holes from theoretical laboratories to real ``particle detectors''
has never been nearer than today.
However paradoxical this may seem, black holes could help us in finding 
one of the most theoretically motivated, but experimentally elusive, particle: The axion.
This is because a rotating black hole can host an axion cloud -- fed by superradiant instability at the expense of the black hole rotational energy -- surrounding it.
 Up to now the properties of such system were studied 
 only considering gravitational interactions.
This is a limitation since any boson with the same mass, irrespective of its particle physics origin, displays the same superradiant physics as long as gravitational interactions are concerned.
 
In this paper we investigated the possible consequences of the parity-violating coupling of the axion with an electromagnetic field in the context of black hole superradiance. 
The key idea is that the axion cloud surrounding a Kerr black hole behaves like an optically active medium, and a ray of light experiences a polarization-dependent bending traveling through it.
Motivated by this picture, we computed the polarization-dependent lensing caused by this phenomenon 
considering the QCD axion, the photo-philic QCD axion, and a generic ALP. 
We discussed the experimental setup that is needed to detect such effect, focusing on 
 the radio observation of a linearly polarized astrophysical source like an AGN.
 We argued that 
 a VLBI array of radio telescopes has the capability to detect the polarization-dependent bending effect caused by the axion cloud surrounding a Kerr black hole, 
 and we delimited the parameter space in which this is relevant in conjunction with other experimental axion searches.

\section*{Acknowledgments}
We thank Diego Blas, Chris Done, Carlos Herdeiro, Jun Hou, Luca Di Luzio and Sergey Sibiryakov
 for discussions, and Paolo Pani for valuable comments on a preliminary version of this manuscript.

\appendix

\section{Radial eigenfunctions and rotating axion cloud}\label{app:Numerics}

The radial eq.~(\ref{eq:KGTortoise}) admits two well-defined limits in the near- and far-horizon region. 
In the far-horizon region, defined by the condition $r \gg M$, $\Delta \simeq r^2(1-2M/r)$,  the radial equation reduces to
\begin{equation}\label{eq:FarRadial}
\frac{d^2(\tilde{r}R_{\rm far})}{d\tilde{r}^2} + \left[
-\frac{1}{4} + \frac{l+n+1}{\tilde{r}} - \frac{l(l+1)}{\tilde{r}^2}
\right]\tilde{r}R_{\rm far} = 0~,
\end{equation}
with $R_{\rm far}$ function of $\tilde{r}$ defined accordingly to eq.~(\ref{eq:RadialApprox}). 
This is the same equation describing an electron in the hydrogen atom, thus enforcing the analogy with Quantum Mechanics. Eq.~(\ref{eq:FarRadial}) can be solved in terms of confluent hypergeometric function
\begin{equation}\label{eq:RadHyper}
R_{\rm far}(\tilde{r}) = \tilde{r}^l e^{-\tilde{r}/2} {_1}F_1(l+1 - \nu;\, 2l+2;\, \tilde{r})~,
\end{equation}
with $\nu = l + n + 1$ the principal quantum number.
The confluent hypergeometric function is given in terms of the Laguerre polynomial by 
\begin{equation}
{\rm L}_n^m(x) = \frac{(m+n)!}{m! n!}{_1}F_1(-n;\,m+1;\,x)~,
\end{equation}
and eq.~(\ref{eq:RadHyper}) reproduces the radial function used in eq.~(\ref{eq:RadialApprox}) that is, therefore, strictly valid only in the far-horizon limit. 
In the near-horizon region, defined by $0< r - r_+ \ll (l/M\mu)^2 M$, the radial equation is solved 
by~\cite{Arvanitaki:2010sy}
\begin{equation}
R_{\rm near}(r) =\left(\frac{r-r_+}{r-r_-}\right)^{-iP} {_2}F_1\left(-l;\,l+1;\,1+2iP;\,\frac{r-r_-}{r_+-r_-}\right)~,~~~~ P\equiv 2r_+\left(
\frac{\omega - m\Omega_+}{r_+ - r_-}
\right)~,
\end{equation} 
where the angular velocity of the black hole horizon is $\Omega_H = \tilde{a}/2r_+$.

 The eigenvalue problem for the radial equation can be solved by means of the continued fraction method championed in~\cite{Leaver:1985ax} (see also~\cite{Dolan:2007mj}, and~\cite{Pani:2013pma} for a pedagogical review about modern black hole perturbation theory). In a nutshell, we look for a radial solution of the form
\begin{equation}\label{eq:Fullradial}
R(r) = \left(
r- r_+
\right)^{-i\sigma}
\left(
r -r_- 
\right)^{i\sigma+\chi -1}
e^{-r\sqrt{\mu^2 - \omega^2}}
\sum_{n=0}^{\infty}a_n
\left(\frac{r- r_+ }{r - r_-}\right)^n~,
\end{equation}
with 
\begin{equation}
\sigma = \frac{2M r_+}{r_+ - r_-}(\omega - m\Omega_H)~,~~~~~~\chi = \frac{M(2\omega^2 - \mu^2)}{\sqrt{\mu^2 - \omega^2}}~.
\end{equation}
Note that this ansatz correctly describes the characteristic asymptotic behavior of  bound states.
Using this expression for $R(r)$, the radial equation 
returns a three-term recurrence relation for
the coefficients $a_n$ that can be solved only for particular values of $\omega = \omega_R + i\omega_I$.
 These are the eigenfrequencies describing bound states. 
 \begin{figure}[!htb!]
\minipage{0.48\textwidth}
  \includegraphics[width=1\linewidth]{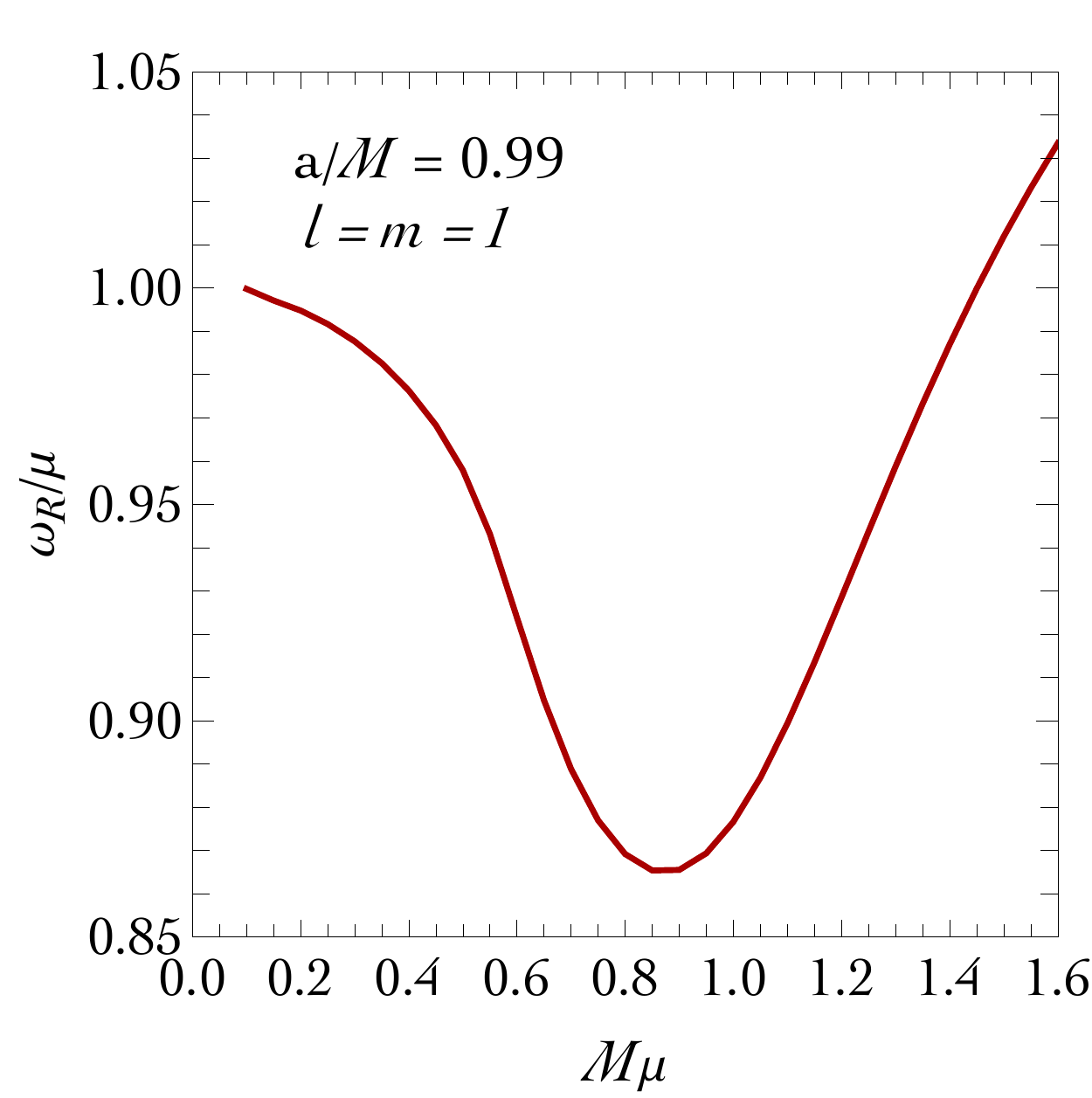}
\endminipage 
\minipage{0.5\textwidth}
  \includegraphics[width=1\linewidth]{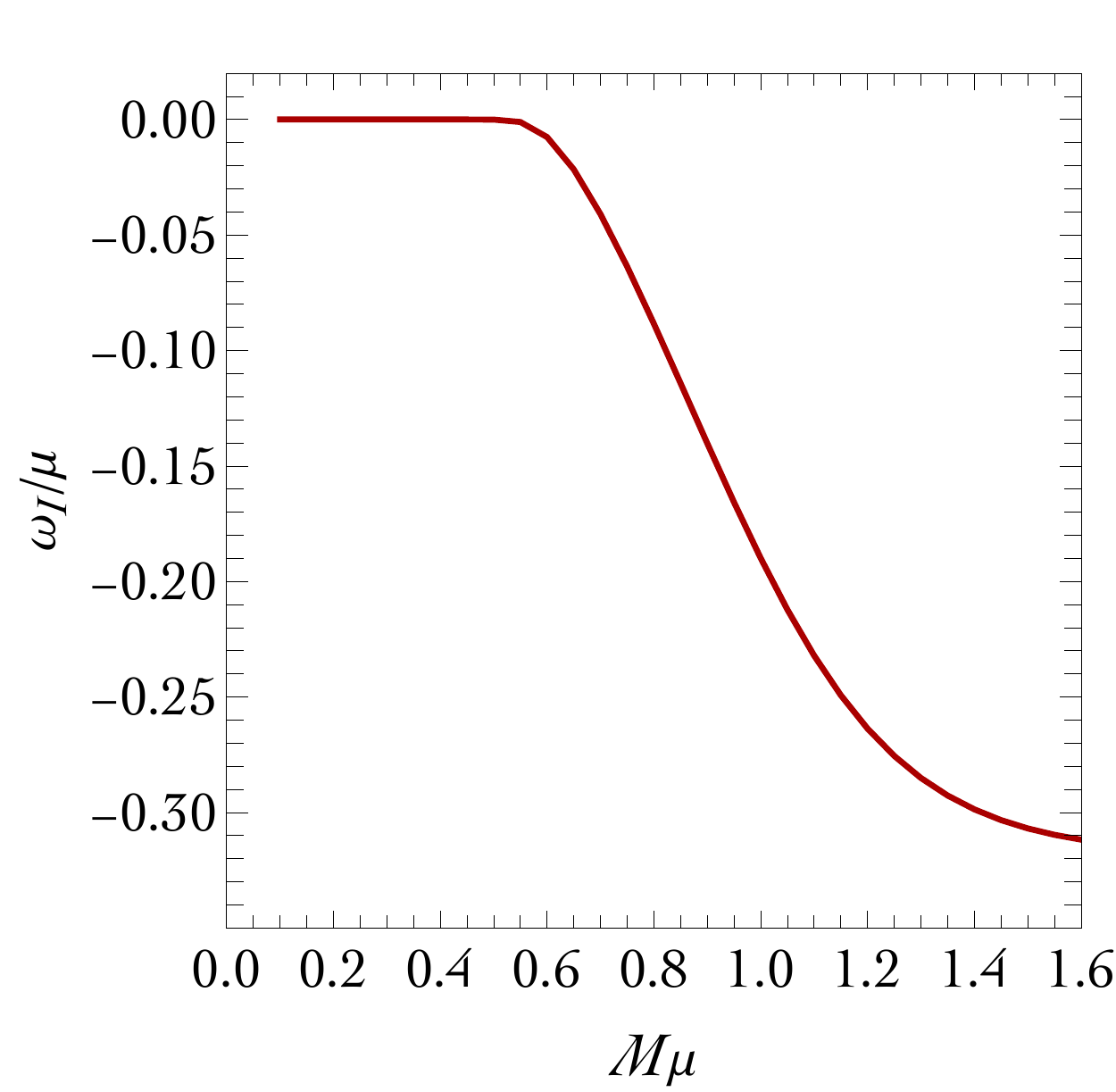}
\endminipage 
\vspace{-.2 cm}
\caption{\label{fig:Omega}\em 
Real and imaginary part (left and right panel, respectively)
of the bound state frequencies for a scalar field in a Kerr background, as a function 
of the dimensionless parameter $M\mu$. We fix the spin parameter $a/M = 0.99$, and we focus on the eigenmode with $l = m = 1$. 
We solved numerically eq.~(\ref{eq:KGTortoise}), and we used the  Leaver's method to obtain the bound 
state frequencies when $M\mu \sim 1$~\cite{Dolan:2007mj}.
}
\end{figure}
We implement numerically the continued fraction method, and 
we show in fig.~\ref{fig:Omega} the values of $\omega_R$ (left panel) and $\omega_I$ (right panel)
obtained by solving the eigenvalue problem for the radial equation.
 In the small $M\mu$ limit,
  the agreement with the approximation 
 used in eqs.~(\ref{eq:LambdaApprox1},\,\ref{eq:LambdaApprox2}) is evident. 
 \begin{figure}[!htb!]
\minipage{0.46\textwidth}
  \includegraphics[width=1\linewidth]{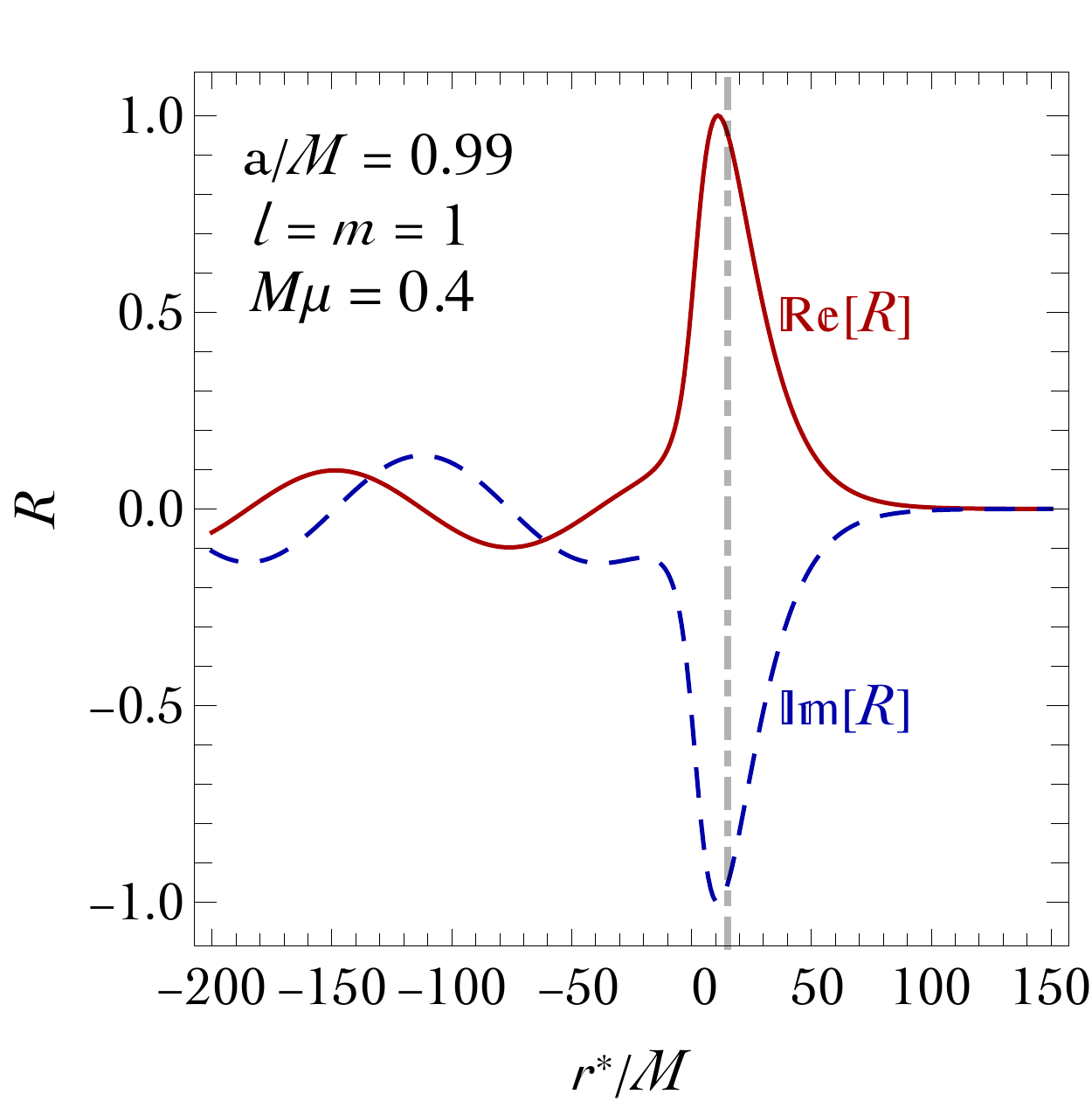}
\endminipage 
\minipage{0.52\textwidth}
  \includegraphics[width=1\linewidth]{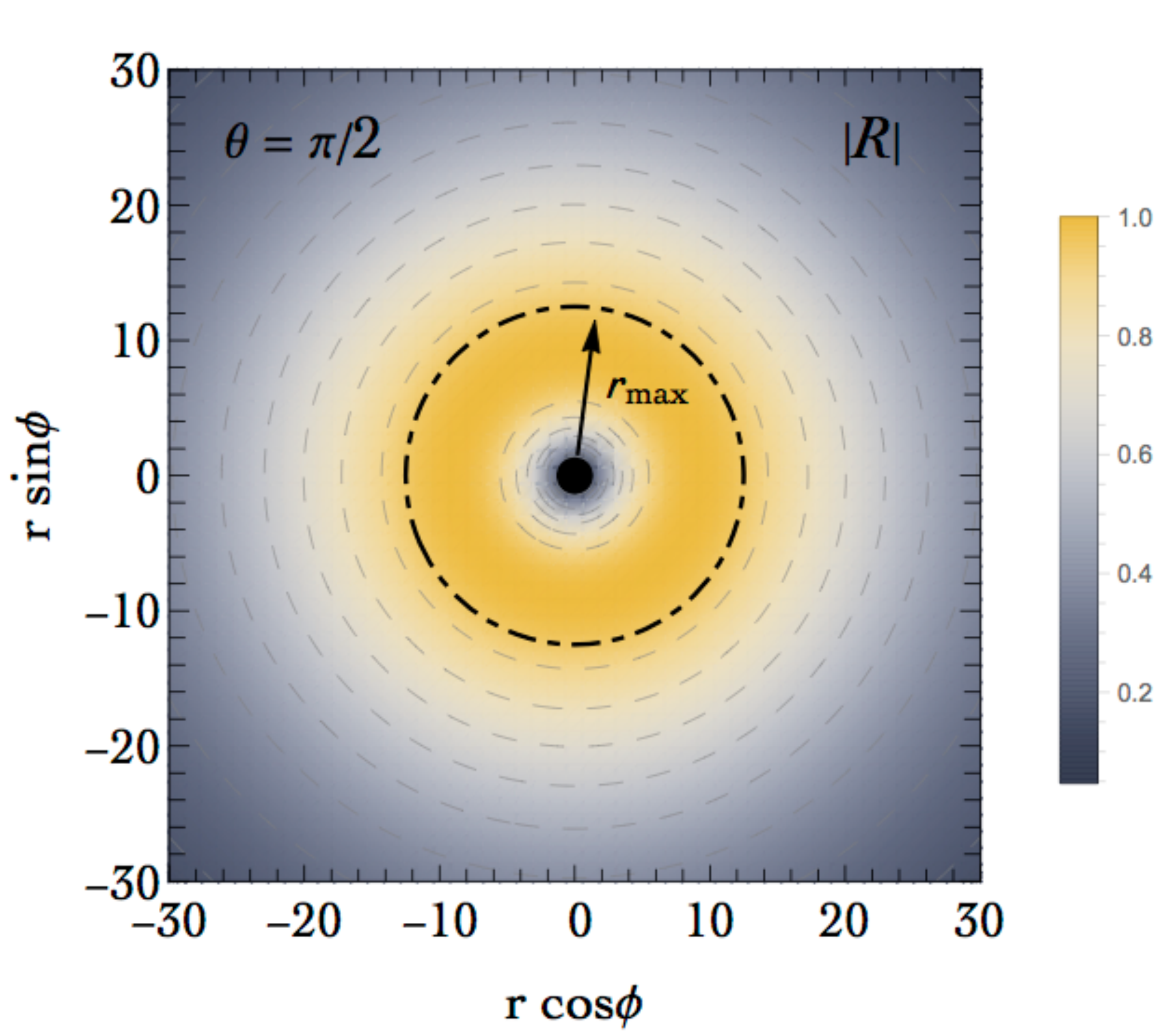}
\endminipage 
\vspace{-.2 cm}
\caption{\label{fig:RadialR}\em 
Left panel. Real (red, solid line) and imaginary (blue, dashed line) part of the radial eigenfunction  $R$ with $l = m = 1$ as a function of the tortoise coordinate $r^*$ 
obtained numerically using the Leaver's method~\cite{Dolan:2007mj}.
For comparison, the vertical gray dot-dashed line indicates at $r^*/M \simeq 15.2$ indicates the position of $\tilde{r}_{\max} = 2$ in terms of the 
tortoise coordinate. Right panel. Density plot of the absolute value $|R|$ (arbitrarily normalized to $1$ at the maximum) in the equatorial plane $\theta = \pi/2$.
The black dot-dashed circle indicates the location of $\tilde{r}_{\max} = 2$ obtained using the analytical approximation in eq.~(\ref{eq:RadialApprox}).
}
\end{figure}
Having computed the bound state frequencies, the full radial eigenfunction can be obtained 
 from eq.~(\ref{eq:Fullradial}). 
 We show our numerical solution in fig.~\ref{fig:RadialR}, and we comment about the 
 comparison with the far-horizon approximation (see caption for details).
\begin{figure}[!htb!]
\begin{center}
  \includegraphics[width=.65\linewidth]{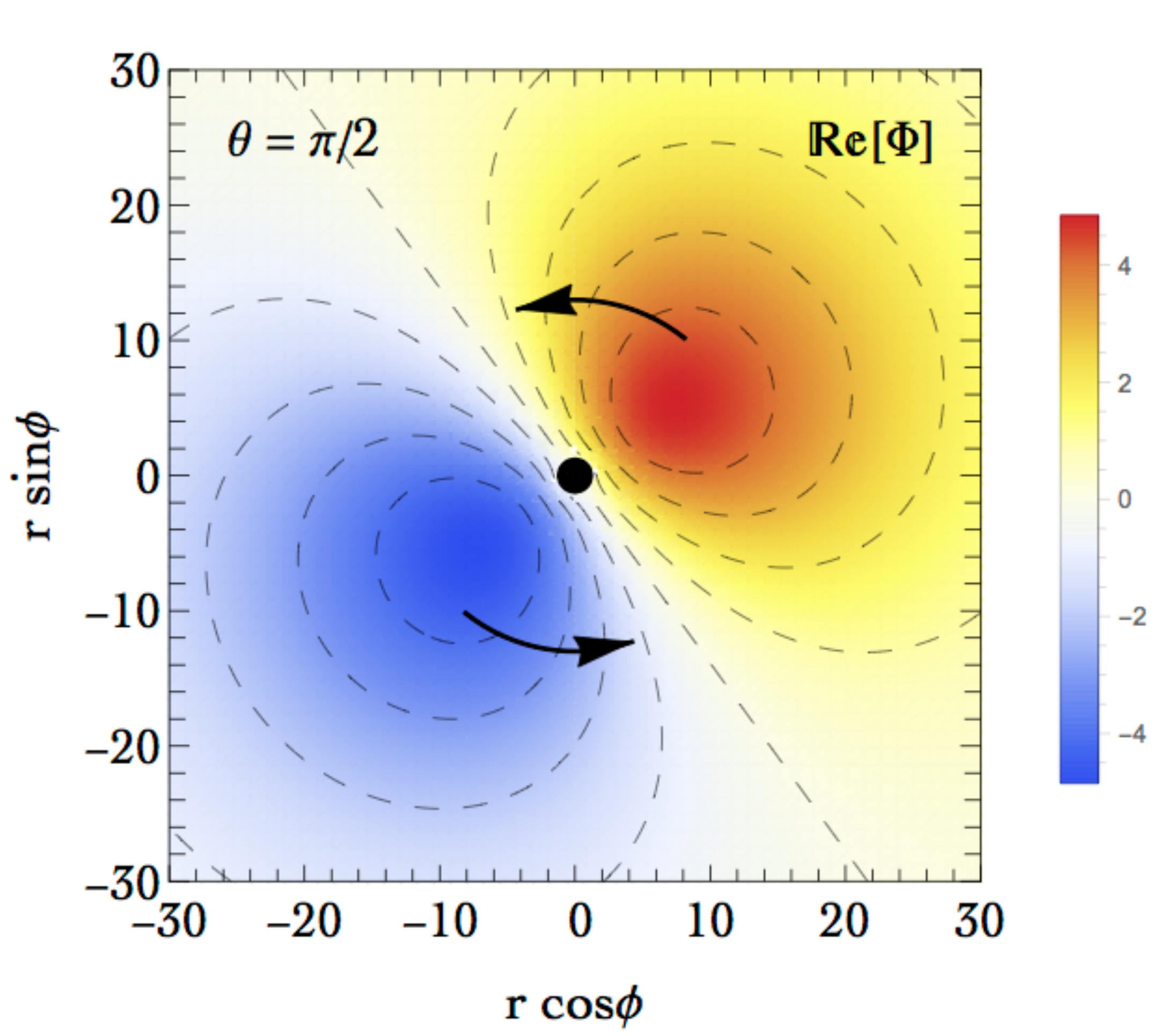}
\end{center}
\vspace{-.25 cm}
\caption{\label{fig:AxionCloud2}\em 
Density plot of the axion cloud $\Re[\Phi] = \Re\left[
e^{im\phi}S_{lm}(\theta)e^{-i\omega t}R_{nl}(r)
\right]$ with $n = 0$, $l = m = 1$ in the equatorial plane. We consider the explicit case with $a/M = 0.99$, $M\mu = 0.4$, and we take for reference $t = 0$.
As time passes by, the axion cloud rotates anti-clockwise in the direction of the black arrows.
The period is $T = 2\pi/\omega_R$.
}
\end{figure}

Finally, it is possible to reconstruct
 the full solution of the Klein-Gordon equation in eq.~(\ref{eq:FullKG}) by including the angular- and time-dependent part.
 For completeness, we show the full solution in the equatorial plane in fig.~\ref{fig:AxionCloud2} (see caption for details).

\section{Modified dispersion relation}\label{app:Dispersion}

In this appendix we derive the dispersion relation in eq.~(\ref{eq:Dispersion}).
From the Lagrangian density 
\begin{equation}
\mathcal{L} = -\frac{1}{4}F_{\mu\nu}F^{\mu\nu} - \frac{g_{a\gamma\gamma}}{2}\left(
\partial_{\mu}\Phi
\right)A_{\nu}\tilde{F}^{\mu\nu}~, 
\end{equation}
we extract the Euler-Lagrange equations of motion $\left[
g^{\mu\nu}\Box - g_{a\gamma\gamma}
\epsilon^{\mu\nu\alpha\beta}
(\partial_{\alpha}\Phi)\partial_{\beta}
\right]A_{\nu}(x) = 0$
which, in Fourier space, give
\begin{equation}\label{eq:EOM}
\left[
g^{\mu\nu}k^2 + ig_{a\gamma\gamma}\epsilon^{\mu\nu\alpha\beta}(\partial_{\alpha}\Phi)k_{\beta}
\right]\tilde{A}_{\nu}(k) \equiv K^{\mu\nu}\tilde{A}_{\nu}(k) = 0~.
\end{equation}
In eq.~(\ref{eq:EOM}) we neglected the second derivative term proportional to $g_{a\gamma\gamma}(\partial_{\mu}\partial_{\rho}\Phi)A_{\sigma}\epsilon^{\rho\sigma\mu\nu}$, in analogy with what discussed in section~\ref{sec:Bending}. We introduce the short-hand notation $\eta_{\alpha}\equiv 
g_{a\gamma\gamma}(\partial_{\alpha}\Phi)$. In order to solve eq.~(\ref{eq:EOM}) we define the operator 
 $S^{\mu}_{~\nu} \equiv \epsilon^{\lambda\mu\alpha\beta}\eta_{\alpha}k_{\beta}\epsilon_{\lambda\nu\rho\sigma}\eta^{\rho}k^{\sigma}$.
The Levi-Civita contraction property
 \begin{equation}
\epsilon_{i_1,\dots,i_k,i_{k+1},\dots,i_n}\epsilon^{i_1,\dots,i_k,j_{k+1},\dots,j_n} = (-1)\,k!\,\delta_{i_{k+1},\dots,i_n}^{j_{k+1},\dots,j_n}~,~~
~~~~~{\rm with}~~~
\delta_{\nu_1,\dots,\nu_p}^{\mu_1,\dots,\mu_p} \equiv\left|
\begin{array}{ccc}
\delta_{\nu_1}^{\mu_1}  & \dots  & \delta_{\nu_p}^{\mu_1}  \\
 \vdots & \ddots  &  \vdots \\
\delta_{\nu_1}^{\mu_p}  & \dots   &   \delta_{\nu_p}^{\mu_p}
\end{array}
\right|~,
\end{equation}
gives the explicit expression
\begin{equation}
S^{\mu\nu} = g^{\mu\nu}\left[
(\eta\cdot k)^2 - \eta^2 k^2
\right] - \eta\cdot k\left(
\eta^{\mu}k^{\nu} + \eta^{\nu}k^{\mu}
\right) + k^2\eta^{\mu}\eta^{\nu} +\eta^2 k^{\mu}k^{\nu}~,
\end{equation}
with the following properties 
\begin{equation}
S^{\mu\nu}k_{\nu} = S^{\mu\nu}\eta_{\nu} = 0~,~~~~S\equiv S^{\mu}_{~\mu} = 2\left[
(\eta\cdot k)^2 - \eta^2 k^2 
\right]~,~~~~S^{\mu\nu}S_{\nu\lambda} = \frac{S}{2}S^{\mu}_{~\lambda}~.
\end{equation}
We can define the two projectors
\begin{equation}
\mathcal{P}_{\pm}^{\mu\nu}\equiv \frac{S^{\mu\nu}}{S} \mp \frac{i}{\sqrt{2S}}\epsilon^{\mu\nu\alpha\beta}
\eta_{\alpha}k_{\beta}~.
\end{equation}
This is a good definition, since we have the following properties
\begin{equation}
\mathcal{P}_{\pm}^{\mu\lambda} \mathcal{P}_{\pm\,\lambda\nu} = \mathcal{P}_{\pm\,\nu}^{~\mu}~,~~~~
\mathcal{P}_{\pm}^{\mu\lambda} \mathcal{P}_{\mp\,\lambda\nu} = 0~.
\end{equation}
Furthermore, $\mathcal{P}_{\pm}^{\mu\nu}k_{\nu}=\mathcal{P}_{\pm}^{\mu\nu}\eta_{\nu} = 0$, 
$g_{\mu\nu}\mathcal{P}_{\pm}^{\mu\nu} = 1$, and $\mathcal{P}_{+}^{\mu\nu}+
\mathcal{P}_{-}^{\mu\nu} = 2S^{\mu\nu}/S$. 
The operator in eq.~(\ref{eq:EOM}) becomes
\begin{equation}
K^{\mu\nu} = g^{\mu\nu}k^2 + \sqrt{\frac{S}{2}}\left(
\mathcal{P}_{-}^{\mu\nu} - \mathcal{P}_{+}^{\mu\nu}
\right)~.
\end{equation}
We now have all the ingredients to derive a dispersion relation from eq.~(\ref{eq:EOM}).  
We start from a space-like unit vector, for example $\varepsilon = (0,i,1,0)/\sqrt{2}$.
We then define the two
 projections $\tilde{\varepsilon}^{\mu}_{\pm}\equiv \mathcal{P}_{\pm}^{\mu\nu}\varepsilon_{\nu}$.
 From the properties of the projectors it follows that 
 \begin{equation}
K^{\mu\nu}\tilde{\varepsilon}_{\pm\,\nu} =  \left[
k^2 \mp \sqrt{\frac{S}{2}}
 \right]\tilde{\varepsilon}_{\pm}^{\mu}~.
 \end{equation}
 Therefore, $\tilde{A}^{\mu} = \tilde{\varepsilon}_{\pm}^{\mu}$ is a solution of eq.~(\ref{eq:EOM}) 
 if and only if $k^2 = \pm \sqrt{S/2}$, or
 \begin{equation}
 k^4 + \eta^2 k^2 = (\eta\cdot k)^2~,
 \end{equation}
 that is the modified dispersion relation presented in eq.~(\ref{eq:Dispersion}). 
 Since the limit $g_{a\gamma\gamma} \to 0$ should recover the standard parity-invariant propagation
  in which there is no difference in the physical properties of a right- and a left-handed 
circularly  polarized electromagnetic wave, it is natural to identify the two distinct solutions 
arising in the case $g_{a\gamma\gamma} \neq 0$ as the two different circular polarizations.

\section{Equation for the photon orbit}\label{app:C}

Let us start from eq.~(\ref{eq:FlatBanding}) in 
 Schwarzschild background
\begin{equation}\label{eq:RadialVel}
\left(
\frac{dr}{d\xi}
\right)^2 = E_{\gamma}^2 - \frac{L^2}{r^2}
\left(
1 - \frac{2M}{r}
\right)
 \mp g_{a\gamma\gamma}
E_{\gamma}\frac{\partial \Phi}{\partial t}~.
\end{equation}
The equation for the photon orbit is given by
\begin{equation}\label{eq:MainEquationOrbit}
\frac{d\phi}{dr} = \frac{d\phi}{d\xi}\frac{d\xi}{dr} = \pm
\frac{1}{r^2\sqrt{
\frac{E_{\gamma}^2}{L^2}\left(
1 \mp \frac{g_{a\gamma\gamma}}{E_{\gamma}}\frac{\partial \Phi}{\partial t}
\right)
 - \frac{1}{r^2}\left(
1-\frac{2M}{r}
\right)
}}~,
\end{equation}
where the minus (plus) sign corresponds to incoming (outgoing) light rays.

\begin{figure}[!htb!]
\vspace{.5cm}
\begin{center}
  \includegraphics[width=.85\linewidth]{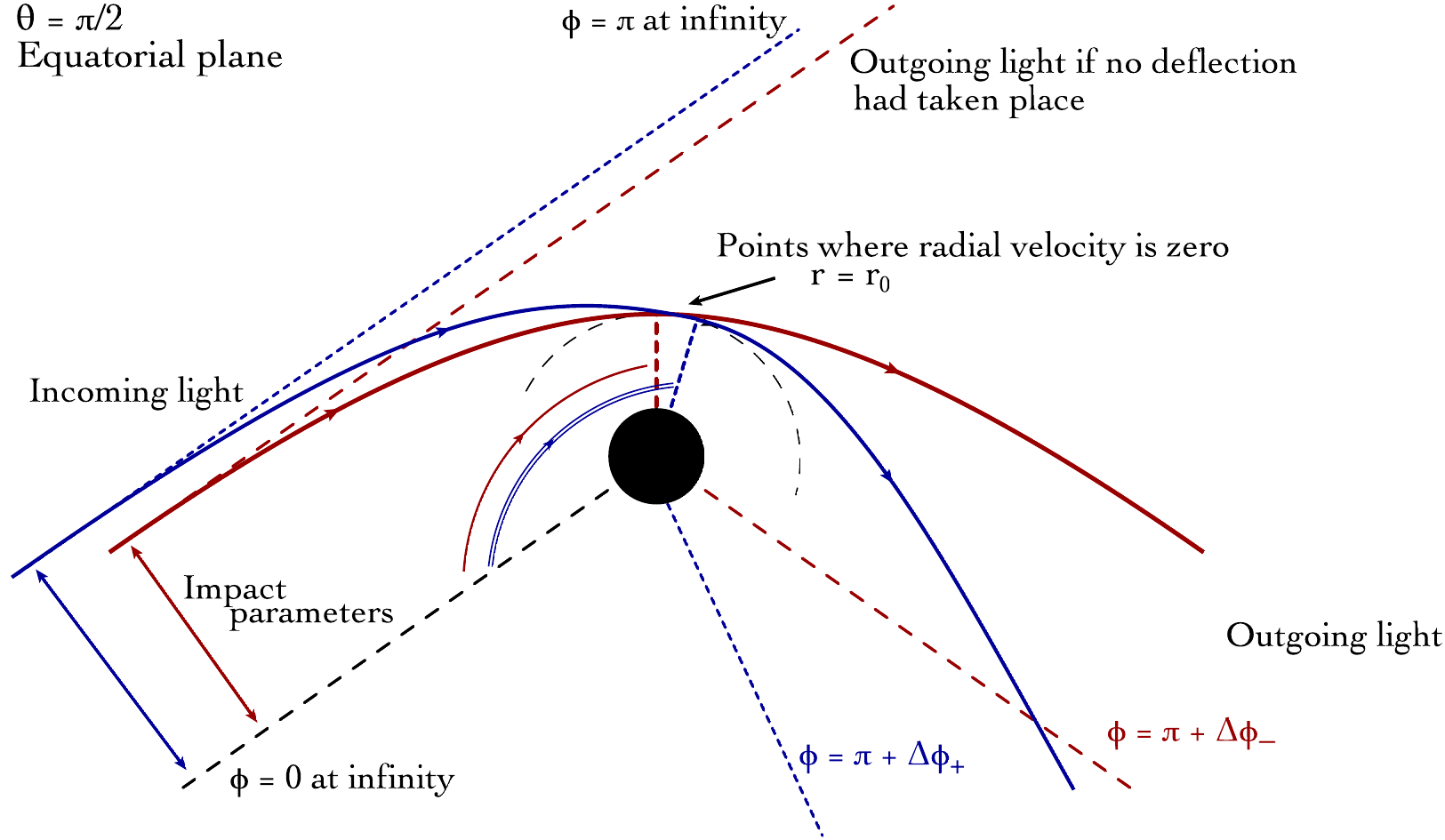}
\end{center}
\vspace{-.25 cm}
\caption{\label{fig:LightBending}\em 
Deflection of a ray of light 
in the gravitational field of a black hole with mass $M$. 
The dashed arc of circumference represents the points at distance $r = r_0$ from the black hole center.
}
\end{figure}

The angle $\phi$ is defined to be $\phi = 0$ for incoming light 
at infinite distance from the black hole. Light traveling in straight line will have $\phi = \pi$ in the opposite outgoing limit. In order to compute the deflection angle we consider the setup illustrated in fig.~\ref{fig:LightBending}.
We follow the standard computation of gravitational lensing.
The distance of closest approach $r_0$ of the light ray is defined by means of the condition $dr/d\xi = 0$.
From eq.~(\ref{eq:RadialVel}) we find
\begin{equation}\label{eq:Impact}
\frac{E_{\gamma}^2}{L^2} = \frac{1-2M/r_0}{r_0^2\left[
1\mp a(E_{\gamma},r_0,\frac{\pi + \Delta\phi_{\mp}}{2})
\right]}~,~~~~~~~~~{\rm with}~~~a(E_{\gamma},r,\phi) \equiv \frac{g_{a\gamma\gamma}}{E_{\gamma}}
\left.
\frac{\partial\Phi}{\partial t}\right|_{r, \phi}~.
\end{equation}
The distance of closest approach defines the angles $\Delta\phi_{\pm}$ as one can see from fig.~\ref{fig:LightBending}.
Note that $E_{\gamma}^2/L^2 = 1/b^2$ defines the impact parameter $b$.
If we fix $r_0$ to be the same for both left- and right-handed circularly polarized waves
we have two different values for the impact parameter, as illustrated in fig.~\ref{fig:LightBending}.
Alternatively, one can fix the impact parameter but in this case the distance of closest approach will differ between the two polarizations.
We can now use the condition in eq.~(\ref{eq:Impact}) into eq.~(\ref{eq:MainEquationOrbit}).
For incoming light rays, we find
\begin{equation}\label{eq:IncomingDeflection}
\frac{d\phi}{dr} = -
\frac{1}{r^2\sqrt{\frac{1}{r_0^2}
\frac{\left[
1\mp a(E_{\gamma},r,\phi)
\right]}{\left[
1\mp a(E_{\gamma},r_0,\frac{\pi + \Delta\phi_{\mp}}{2})
\right]}\left(
1-\frac{2M}{r_0}
\right)
 - \frac{1}{r^2}\left(
1-\frac{2M}{r}
\right)
}}~.
\end{equation}
This equation must be integrated between $r = \infty$ and $r = r_0$ in order to obtain the deflection angle 
for incoming light rays. The final deflection angle, 
$\pi + \Delta\phi_{\pm}$, is obtained by adding the corresponding integration -- in the interval between $r = r_0$ and $r = \infty$ -- for outgoing light rays, 
as illustrated in fig.~\ref{fig:LightBending}.

We can use the following approximation in eq.~(\ref{eq:IncomingDeflection}).
In our computation we take the distance of closest approach to be $r_0 = r_{\rm max}$.
Furthermore, we introduce the dimensionless variable $x \equiv r/M$, and we find
\begin{equation}
\frac{d\phi}{dx} = -
\frac{1}{x^2\sqrt{
\frac{1}{x_{\rm max}^2}
\frac{\left[
1\mp a(E_{\gamma},x,\phi)
\right]}{\left[
1\mp a(E_{\gamma},x_{\rm max},\frac{\pi + \Delta\phi_{\mp}}{2})
\right]}\left(
1-\frac{2}{x_{\rm max}}
\right)
 - \frac{1}{x^2}\left(
1-\frac{2}{x}
\right)
}}~.
\end{equation}
The flat space limit is
\begin{equation}
\frac{d\phi}{dx} = -
\frac{1}{x^2\sqrt{
\frac{1}{x_{\rm max}^2}
\frac{\left[
1\mp a(E_{\gamma},x,\phi)
\right]}{\left[
1\mp a(E_{\gamma},x_{\rm max},\frac{\pi + \Delta\phi_{\mp}}{2})
\right]}
 - \frac{1}{x^2}
}}~.
\end{equation}
Let us now expand the right-hand side for small $a$. We find
\begin{equation}
\frac{d\phi}{dx} = - \frac{1}{x^2\sqrt{\frac{1}{x_{\rm max}^2}  - \frac{1}{x^2}}} 
\mp
\frac{a(E_{\gamma},x,\phi) - a(E_{\gamma},x_{\rm max},\frac{\pi + \Delta\phi_{\mp}}{2})}
{2x^2 x_{\rm max}^2\left(\frac{1}{x_{\rm max}^2} - \frac{1}{x^2}\right)^{3/2}}~.
\end{equation}
The first term reproduces the trivial flat space limit, and the integration 
between $x = \infty$ and $x = x_{\rm max}$ gives the angle $\phi = \pi/2$ corresponding to 
outgoing light with no deflection, as illustrated in fig.~\ref{fig:LightBending}. 
Since by definition $\Delta\phi_{\pm} \sim \mathcal{O}(g_{a\gamma\gamma})$, at the first order in the coupling $g_{a\gamma\gamma}$ we can write 
\begin{equation}
\frac{d\phi}{dx} = - \frac{1}{x^2\sqrt{\frac{1}{x_{\rm max}^2}  - \frac{1}{x^2}}} 
\mp
\frac{a(E_{\gamma},x,\phi) - a(E_{\gamma},x_{\rm max},\frac{\pi}{2})}
{2x^2 x_{\rm max}^2\left(\frac{1}{x_{\rm max}^2} - \frac{1}{x^2}\right)^{3/2}}~,
\end{equation}
that is the equation for the photon orbit that we solved in section~\ref{sec:Bending}.





\newpage

\def\hhref#1{\href{http://arxiv.org/abs/#1}{#1}} 

\end{document}